\documentclass[12pt,eqsecnum]{revtex4}

\usepackage{amssymb}
\usepackage{amsmath}
\usepackage{graphicx}
\usepackage{amsbsy}

\newcommand{\bq}{\begin{eqnarray}}
\newcommand{\eq}{\end{eqnarray}}
\newcommand{\bqn}{\begin{eqnarray*}}
\newcommand{\eqn}{\end{eqnarray*}}
\newcommand{\rr}{{\bf r}}

\newcommand{\qq}{{\bf q}}

\newcommand{\erf}{\mathop\text{erf}}
\newcommand{\erfc}{\mathop\text{erfc}}
\DeclareMathAlphabet{\mathpzc}{OT1}{pzc}{m}{it}
\newlength{\GraphicsWidth}
\newcommand{\hx}{\hat{x}}

\begin{document}
\title{Two-dimensional one-component plasma \\ 
on a Flamm's paraboloid} 
\author{Riccardo Fantoni}
\affiliation{Istituto Nazionale per la Fisica della Materia and Dipartimento
di Chimica Fisica, Universit\`{a} di Venezia, S. Marta DD 2137, I-30123
Venezia, Italy}
\author{Gabriel T\'ellez}
\affiliation{Grupo de F\'{\i}sica T\'eorica de la
Materia Condensada, Departamento de F\'{\i}sica, Universidad de Los
Andes, A.A. 4976, Bogot\'a, Colombia}

\begin{abstract}
We study the classical non-relativistic two-dimensional one-component
plasma at Coulomb coupling $\Gamma=2$ on the Riemannian surface known
as Flamm's paraboloid which is obtained from the spatial part of the
Schwarzschild metric. At this special value of the coupling constant,
the statistical mechanics of the system are exactly solvable
analytically. The Helmholtz free energy asymptotic expansion for the
large system has been found. The density of the plasma, in the
thermodynamic limit, has been carefully studied in various situations.
\end{abstract}

\maketitle

\noindent Keywords: Coulomb systems, one-component plasma, non
constant curvature.

\section{Introduction}

The system under consideration is a classical (non quantum)
two-dimensional one-component plasma: a system composed of one species
of charged particles living in a two-dimensional surface, immersed in
a neutralizing background, and interacting with the Coulomb potential.
The one-component classical Coulomb plasma is exactly solvable in one
dimension \cite{Edwards62}.  In two dimensions, in their 1981 work,
B.~Jancovici and A.~Alastuey~\cite{Jancovici81b,Alastuey81} showed how
the partition function and $n$-body correlation functions of the
two-dimensional one-component classical Coulomb plasma (2dOCP) on a
plane can be calculated exactly analytically at the special value of
the coupling constant $\Gamma=\beta q^2=2$, where $\beta$ is the
inverse temperature and $q$ the charge carried by the particles. This
has been a very important result in statistical physics since there
are very few analytically solvable models of continuous fluids in
dimensions greater than one.

Since then, a growing interest in two-dimensional plasmas has lead to
study this system on various flat geometries
\cite{Rosinberg84,Jancovici94,Jancovici96} and two-dimensional curved
surfaces: the cylinder~\cite{Choquard81,Choquard83}, the sphere
\cite{Caillol81,Jancovici92,Jancovici96b,Tellez99,Jancovici00} and the
pseudosphere~\cite{Jancovici98,Fantoni03jsp,Jancovici04}. These
surface have constant curvature and the plasma there is
homogeneous. Therefore, it is interesting to study a case where the
surface does not have a constant curvature.

In this work we study the 2dOCP on the Riemannian surface ${\cal S}$
known as the Flamm's paraboloid, which is obtained from the spatial
part of the Schwarzschild metric. The Schwarzschild geometry in
general relativity is a vacuum solution to the Einstein field
equation which is spherically symmetric and in a two dimensional world
its spatial part has the form 
\begin{eqnarray} \label{metric}
d\mathbf{s}^2=\left(1-\frac{2M}{r}\right)^{-1}\,dr^2+r^2\,d\varphi^2~.
\end{eqnarray} 
In general relativity, $M$ (in appropriate units) is the mass of the
source of the gravitational field. This surface has a hole of radius
$2M$ and as the hole shrinks to a point (limit $M\to 0$) the surface
becomes flat. It is worthwhile to stress that, while the Flamm's
paraboloid considered here naturally arises in general relativity, we
will study the classical ({\sl i.e.\/} non quantum) statistical
mechanics of the plasma obeying non-relativistic dynamics. Recent
developments for a statistical physics theory in special relativity
have been made in~\cite{kaniadakis02,kaniadakis05}. To the best of our
knowledge no attempts have been made to develop a statistical
mechanics in the framework of general relativity.

The ``Schwarzschild wormhole'' provides a path from the upper
``universe'' to the lower one. We will study the 2dOCP on a single
universe, on the whole surface, and on a single universe with the
``horizon'' (the region $r=2M$) grounded.

Since the curvature of the surface is not a constant but varies from
point to point, the plasma will not be uniform even in the
thermodynamic limit.

We will show how the Coulomb potential between two unit charges on
this surface is given by $-\ln(|z_1-z_2|/\sqrt{|z_1z_2|})$ where
$z_i=(\sqrt{r_i}+\sqrt{r_i-2M})^2e^{i\varphi_i}$. This simple form
will allow us to determine analytically the partition function and the
$n$-body correlation functions at $\Gamma=2$ by extending the original
method of Jancovici and Alastuey~\cite{Jancovici81b, Alastuey81}.  We
will also compute the thermodynamic limit of the free energy of the
system, and its finite-size corrections. These finite-size corrections
to the free energy will contain the signature that Coulomb systems can
be seen as critical systems in the sense explained
in~\cite{Jancovici94,Jancovici96}.

The work is organized as follows: in section \ref{sec:model}, we
describe the one-component plasma model and the Flamm's paraboloid,
{\sl i.e.\/} the Riemannian surface ${\cal S}$ where the plasma is
embedded. In section \ref{sec:poisson}, we find the Coulomb pair
potential on the surface ${\cal S}$ and the particle-background
potential. We found it convenient to split this task into three
cases. We first solve Poisson equation on just the upper half of the
surface ${\cal S}$. We then find the solution on the whole surface and
at last we determine the solution in the grounded horizon case. In
section \ref{sec:correlations}, we determine the exact analytical
expression for the partition function and density at $\Gamma=2$ for
the 2dOCP on just one half of the surface, on the whole surface, and
on the surface with the horizon grounded. In section
\ref{sec:conclusions}, we outline the conclusions.

\section{The model}
\label{sec:model}

A one-component plasma is a system of $N$ pointwise particles of
charge $q$ and density $n$ immersed in a neutralizing background
described by a static uniform charge distribution of charge density
$\rho_b=-qn_b$.

In this work, we want to study a two-dimensional one-component plasma
(2dOCP) on a Riemannian surface ${\cal S}$ with the following metric
\begin{eqnarray} 
d\mathbf{s}^2=g_{\mu\nu}dx^\mu dx^\nu=
\left(1-\frac{2M}{r}\right)^{-1}dr^2+r^2d\varphi^2~.
\end{eqnarray}
or $g_{rr}=1/(1-2M/r),g_{\varphi\varphi}=r^2$, and $g_{r\varphi}=0$.

\begin{figure}
\begin{center}
\includegraphics[width=\GraphicsWidth]{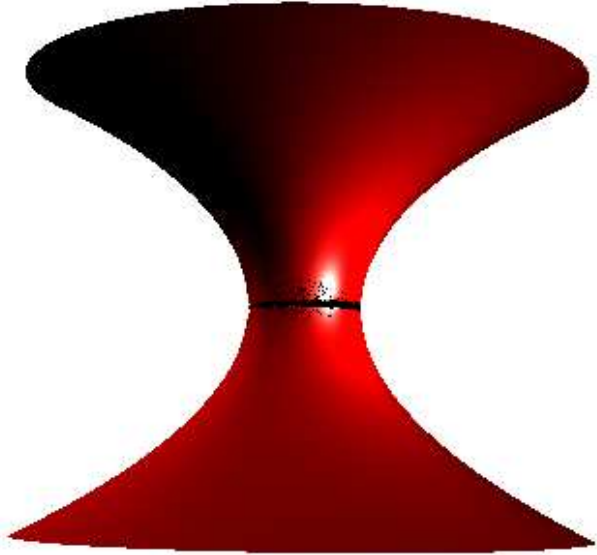}
\end{center}
\caption{The Riemannian surface ${\cal S}$: the Flamm's paraboloid.}
\label{fig:surf}
\end{figure}

This is an embeddable surface in the three-dimensional Euclidean space
with cylindrical coordinates $(r,\varphi,Z)$ with
$d\mathbf{s}^2=dZ^2+dr^2+r^2d\varphi^2$, whose equation is
\begin{eqnarray} \label{surf} Z(r)=\pm 2\sqrt{2M(r-2M)}~.  
\end{eqnarray} 
This surface is illustrated in Fig.~\ref{fig:surf}. It has a hole of
radius $2M$. We will from now on call the $r=2M$ region of the surface
its ``horizon''.

\subsubsection{The Flamm's paraboloid $\cal S$}
\label{sec:topology}

The surface ${\cal S}$ whose local geometry is fixed by the metric
(\ref{metric}) is known as the Flamm's paraboloid. It is composed by
two identical ``universes'': ${\cal S}_+$ the one at $Z>0$, and ${\cal
  S}_-$ the one at $Z<0$. These are both multiply connected surfaces
with the ``Schwarzschild wormhole'' providing the path from one to the
other.

The system of coordinates $(r,\varphi)$ with the metric~(\ref{metric})
has the disadvantage that it requires two charts to cover the whole
surface ${\cal S}$. It can be more convenient to use the variable
\begin{eqnarray} \label{u}
u=\frac{Z}{4M}=
\pm\sqrt{\frac{r}{2M}-1}
\end{eqnarray}
instead of $r$. Replacing $r$ as a function of $Z$ using
equation~(\ref{surf}) gives the following metric when using the system
of coordinates $(u,\varphi)$,
\begin{eqnarray}
d\mathbf{s}^2=4M^2(1+u^2)\left[4\,du^2+(1+u^2)\,d\varphi^2\right]~.
\end{eqnarray}
The region $u>0$ corresponds to ${\cal S}_+$ and the region $u<0$
to ${\cal S}_{-}$.

Let us consider that the OCP is confined in a ``disk'' defined as
\begin{eqnarray}
\Omega_R^{+}=\{\qq=(r,\varphi)\in {\cal S}_+ |0\le\varphi\le 2\pi, 
2M\le r\le R\}~.
\end{eqnarray}
The area of this disk is given by
\begin{eqnarray} \label{volume}
\mathcal{A}_R=\int_{\Omega_R} dS=\pi\left[\sqrt{R(R-2M)}(3M+R)+
6M^2\ln\left(\frac{\sqrt{R}+\sqrt{R-2M}}{\sqrt{2M}}\right)\right]~,
\end{eqnarray}
where $dS=\sqrt{g}\,dr\,d\varphi$ and $g=\det(g_{\mu\nu})$. The
perimeter is $\mathcal{C}_R=2\pi R$.

The Riemann tensor in a two dimensional space has only
$2^2(2^2-1)/12=1$ independent component. In our case the
characteristic component is 
\begin{eqnarray} 
  {R^r}_{\varphi r\varphi}=-\frac{M}{r}~.
\end{eqnarray} 
The scalar curvature is then given by the following indexes
contractions
\begin{eqnarray}
\mathcal{R}={R^\mu}_\mu={R^{\mu\nu}}_{\mu\nu}=2{R^{r\varphi}}_{r\varphi}
=2g^{\varphi\varphi}{R^r}_{\varphi r\varphi}= -\frac{2M}{r^3}~, 
\end{eqnarray}
and the (intrinsic) Gaussian curvature is
$K=\mathcal{R}/2=-M/r^3$. The (extrinsic) mean curvature of the
manifold turns out to be $H=-\sqrt{M/8r^3}$.

The Euler characteristic of the disk $\Omega_R^{+}$ is given by 
\begin{eqnarray}
\chi=\frac{1}{2\pi}\left(\int_{\Omega_R^{+}}K\,dS+
\int_{\partial\Omega_R^{+}}k\,dl\right)~, 
\end{eqnarray} where $k$ is
the geodesic curvature of the boundary $\partial\Omega_R^{+}$.  The
Euler characteristic turns out to be zero, in agreement with the
Gauss-Bonnet theorem $\chi=2-2h-b$ where $h=0$ is the number of
handles and $b=2$ the number of boundaries.

We can also consider the case where the system is confined in a
``double'' disk
\begin{equation}
\Omega_R=\Omega_R^{+}\cup\Omega_R^{-}\,,  
\end{equation}
with $\Omega_R^{-}=\{\qq=(r,\varphi)\in {\cal S}_{-} |0\le\varphi\le
2\pi, 2M\le r\le R\}$, the disk image of $\Omega_{R}^{+}$ on the lower
universe ${\cal S}_{-}$ portion of ${\cal S}$. The Euler
characteristic of $\Omega_R$ is also $\chi=0$.

\subsubsection{A useful system of coordinates}
\label{sec:good-coordinates}

The Laplacian for a function $f$ is
\begin{eqnarray} \nonumber
\Delta f&=&
\frac{1}{\sqrt{g}}\frac{\partial}{\partial q^\mu}
\left(\sqrt{g}\,g^{\mu\nu}\frac{\partial}{\partial q^\nu}\right)f\\
&=&\left[\left(1-\frac{2M}{r}\right)\frac{\partial^2}{\partial r^2}
+\frac{1}{r^2}\frac{\partial^2}{\partial \varphi^2}
+\left(\frac{1}{r}-\frac{M}{r^2}\right)\frac{\partial}{\partial r}\right]
f~,
\end{eqnarray}
where $\qq\equiv(r,\varphi)$. In 
appendix~\ref{app:green}, we show how, finding the
Green function of the Laplacian, naturally leads to consider the
system of coordinates $(x,\varphi)$, with
\begin{equation}
  x=(\sqrt{u^2+1}+u)^{2}
  \,.
\end{equation}
The range for the variable $x$ is $\left]0,+\infty\right[$. The lower
    paraboloid ${\cal S}_{-}$ corresponds to the region $0<x<1$ and
    the upper one ${\cal S}_{+}$ to the region $x>1$. A point in the
    upper paraboloid with coordinate $(x,\varphi)$ has a mirror image
    by reflection ($u\to -u$) in the lower paraboloid, with
    coordinates $(1/x,\varphi)$, since if
\begin{equation}
  x=(\sqrt{u^2+1}+u)^{2}
\end{equation}
then
\begin{equation}
  \frac{1}{x}=(\sqrt{u^2+1}-u)^{2}
  \,.
\end{equation}
In the upper paraboloid ${\cal S}_{+}$, the new coordinate $x$ can be
expressed in terms of the original one, $r$, as
\begin{equation}
x=\frac{(\sqrt{r}+\sqrt{r-2M})^2}{2M}  
\,.
\end{equation}

Using this system of coordinates, the metric takes the form of a
flat metric multiplied by a conformal factor
\begin{equation}
  \label{eq:metric-in-x}
  d\mathbf{s}^2=
  \frac{M^2}{4}\left(1+\frac{1}{x}\right)^4
  \left(dx^2+x^2\,d\varphi^2\right)\,.
\end{equation}
The Laplacian also takes a simple form
\begin{equation}
  \Delta f =\frac{4}{M^2\left(1+\frac{1}{x}\right)^4}
  \,\Delta_{\mathrm{flat}}f
\end{equation}
where 
\begin{equation}
  \Delta_{\mathrm{flat}}f=
  \frac{\partial^2 f}{\partial x^2}
  +\frac{1}{x}\frac{\partial f}{\partial x}
  +\frac{1}{x^2}\frac{\partial^2 f}{\partial \varphi^2}
\end{equation}
is the Laplacian of the flat Euclidean space $\mathbb{R}^2$.  The
determinant of the metric is now given by $g=[ M^2 x (1+x^{-1})^4
/4]^2$.

With this system of coordinates $(x,\varphi)$, the area of a ``disk''
$\Omega_{R}^{+}$ of radius $R$ [in the original system $(r,\varphi)$]
is given by
\begin{equation}
  \mathcal{A}_R=\frac{\pi M^2}{4}\,p(x_m)
\end{equation}
with
\begin{equation}
  \label{eq:p(x)}
  p(x)=x^2+8 x-\frac{8}{x}-\frac{1}{x^2}+12\ln x
\end{equation}
and $x_m=(\sqrt{R}+\sqrt{R-2M})^2/(2M)$.

\section{Coulomb potential}
\label{sec:poisson}

\subsection{Coulomb potential created by a point charge}
\label{sec:green}

The Coulomb potential $G(x,\varphi;x_0,\varphi_0)$ created at
$(x,\varphi)$ by a unit charge at $(x_0,\varphi_0)$ is given by
the Green function of the Laplacian
\begin{equation}
  \label{eq:LaplaceGreen}
  \Delta G(x,\varphi;x_0,\varphi_0)
  =-2\pi \delta^{(2)}(x,\varphi;x_0,\varphi_0)
\end{equation}
with appropriate boundary conditions. The Dirac distribution is given
by
\begin{equation}
  \delta^{(2)}(x,\varphi;x_0,\varphi_0)
  =\frac{4}{M^2 x (1+x^{-1})^4}\,\delta(x-x_0)\delta(\varphi-\varphi_0)
\end{equation}

Notice that using the system of coordinates $(x,\varphi)$ the
Laplacian Green function equation takes the simple form
\begin{equation}
  \label{eq:GreenLaplace-flat}
  \Delta_{\mathrm{flat}}  G(x,\varphi;x_0,\varphi_0)
    =-2\pi\frac{1}{x}\,\delta(x-x_0)\delta(\varphi-\varphi_0)
\end{equation}
which is formally the same Laplacian Green function equation for 
flat space.

We shall consider three different situations: when the
particles can be in the whole surface ${\cal S}$, or when the
particles are confined to the upper paraboloid universe ${\cal
  S}_{+}$, confined by a hard wall or by a grounded perfect conductor.

\subsubsection{Coulomb potential $G^{\mathrm{ws}}$ when the particles live in the whole surface ${\cal S}$}

To complement the Laplacian Green function
equation~(\ref{eq:LaplaceGreen}), we impose the usual boundary
condition that the electric field $-\nabla G$ vanishes at infinity
($x\to\infty$ or $x\to0$). Also, we require the usual interchange
symmetry $G(x,\varphi;x_0,\varphi_0)=G(x_0,\varphi_0;x,\varphi)$ to be
satisfied. Additionally, due to the symmetry between each universe
${\cal S}_{+}$ and ${\cal S}_{-}$, we require that the Green function
satisfies the symmetry relation
\begin{equation}
  \label{eq:symmetry-S+S-}
  G^{\mathrm{ws}}(x,\varphi;x_0,\varphi_0)=
  G^{\mathrm{ws}}(1/x,\varphi;1/x_0,\varphi_0)
\end{equation}

The Laplacian Green function equation~(\ref{eq:LaplaceGreen}) can be
solved, as usual, by using the decomposition as a Fourier
series. Since equation~(\ref{eq:LaplaceGreen}) reduces to the flat
Laplacian Green function equation~(\ref{eq:GreenLaplace-flat}), the
solution is the standard one
\begin{equation}
  \label{eq:Fourier}
  G(x,\varphi;x_0,\varphi_0)=
  \sum_{n=1}^{\infty}
  \frac{1}{n}\left(\frac{x_{<}}{x_{>}}\right)^{2n}
  \cos\left[ n(\varphi-\varphi_0)\right]
  +g_0(x,x_0)
\end{equation}
where $x_{>}=\max(x,x_0)$ and $x_{<}=\min(x,x_0)$. The Fourier
coefficient for $n=0$, has the form
\begin{equation}
  g_0(x,x_0)=
  \begin{cases}
  a_0^{+}\ln x+b_0^{+}\,,&x>x_0\\    
  a_0^{-}\ln x+b_0^{-}\,,&x<x_0\,.
  \end{cases}
\end{equation}
The coefficients $a_0^{\pm},b_0^{\pm}$ are determined by the boundary
conditions that $g_0$ should be continuous at $x=x_0$, its derivative
discontinuous $\partial_x g_0|_{x=x_0^{+}}-\partial_x
g_0|_{x=x_0^{-}}=-1/x_0$, and the boundary condition at infinity
$\nabla g_0|_{x\to\infty}=0$ and $\nabla g_0|_{x\to 0}=0$.
Unfortunately, the boundary condition at infinity is trivially
satisfied for $g_0$, therefore $g_0$ cannot be determined only with
this condition. In flat space, this is the reason why the Coulomb
potential can have an arbitrary additive constant added to
it. However, in our present case, we have the additional symmetry
relation~(\ref{eq:symmetry-S+S-}) which should be satisfied. This
fixes the Coulomb potential up to an additive constant $b_0$. We find
\begin{equation}
  \label{eq:g0-ws}
  g_0(x,x_0)=-\frac{1}{2}\ln\frac{x_{>}}{x_{<}} + b_0\,,
\end{equation}
and summing explicitly the Fourier series~(\ref{eq:Fourier}), we obtain
\begin{equation}
\label{eq:Gws}
G^{\mathrm{ws}}(x,\varphi;x_0,\varphi_0)=
-\ln\frac{\left|z-z_0\right|}{\sqrt{\left|z z_0  \right|}}
+b_0
~,
\end{equation}
where we defined $z=xe^{i\varphi}$ and $z_0=x_0e^{i\varphi_0}$. Notice
that this potential does not reduce exactly to the flat one when
$M=0$. This is due to the fact that the whole surface $\mathcal{S}$ in
the limit $M\to0$ is not exactly a flat plane $\mathbb{R}^2$, but
rather it is two flat planes connected by a hole at the origin, this
hole modifies the Coulomb potential.

\subsubsection{Coulomb potential $G^{\mathrm{hs}}$ when the particles
  live in the half surface ${\cal S}_{+}$ confined by hard walls}

We consider now the case when the particles are restricted to live in
the half surface ${\cal S}_{+}$, $x>1$, and they are confined by a
hard wall located at the ``horizon'' $x=1$. The region $x<1$ (${\cal
  S}_{-}$) is empty and has the same dielectric constant as the upper
region occupied by the particles. Since there are no image charges,
the Coulomb potential is the same $G^{\mathrm{ws}}$ as above. However,
we would like to consider here a new model with a slightly different
interaction potential between the particles. Since we are dealing only
with half surface, we can relax the symmetry
condition~(\ref{eq:symmetry-S+S-}). Instead, we would like to consider
a model where the interaction potential reduces to the flat Coulomb
potential in the limit $M\to0$. The solution of the Laplacian Green
function equation is given in Fourier series by
equation~(\ref{eq:Fourier}). The zeroth order Fourier component $g_0$ can
be determined by the requirement that, in the limit $M\to0$, the
solution reduces to the flat Coulomb potential
\begin{equation}
  G^{\mathrm{flat}}(\rr,\rr')=-\ln\frac{|\rr-\rr'|}{L} 
\end{equation}
where $L$ is an arbitrary constant length. Recalling that $x\sim 2r/M$,
when $M\to0$, we find
\begin{equation}
  \label{eq:g0-hs}
  g_0(x,x_0)=-\ln x_{>}-\ln\frac{M}{2L}
\end{equation}
and 
\begin{equation}
  \label{cgreen}
  G^{\mathrm{hs}}
  (x,\varphi;x_0,\varphi_0)=-\ln |z-z_0|-\ln\frac{M}{2L}
  \,.
\end{equation}

\subsubsection{Coulomb potential $G^{\mathrm{gh}}$ when the particles
  live in the half surface ${\cal S}_{+}$ confined by a grounded
  perfect conductor}

Let us consider now that the particles are confined to ${\cal S}_{+}$
by a grounded perfect conductor at $x=1$ which imposes Dirichlet
boundary condition to the electric potential. The Coulomb potential
can easily be found from the Coulomb potential
$G^{\mathrm{ws}}$~(\ref{eq:Gws}) using the method of images
\begin{equation}
  \label{ghgreen}
  G^{\mathrm{gh}}(x,\varphi;x_0,\varphi_0)=
  -\ln\frac{|z-z_0|}{\sqrt{|z z_0|}}
  +\ln\frac{|z-\bar{z}_0^{-1}|}{\sqrt{|z \bar{z}_0^{-1}|}}
  =
  -\ln\left|\frac{z-z_0}{1-z\bar{z}_0}\right|
\end{equation}
where the bar over a complex number indicates its complex
conjugate. We will call this the grounded horizon Green
function. Notice how its shape is the same of the Coulomb potential on
the pseudosphere~\cite{Fantoni03jsp} or in a flat disk confined by
perfect conductor boundaries~\cite{Jancovici96}.

This potential can also be found using the Fourier
decomposition. Since it will be useful in the following, we note that
the zeroth order Fourier component of $G^{\mathrm{gh}}$ is
\begin{equation}
  \label{eq:g0-gh}
  g_0(x,x_0)=\ln x_{<}\,.
\end{equation}

\subsection{The background}

The Coulomb potential generated by the background, with a constant
surface charge density $\rho_b$ satisfies the Poisson equation
\begin{eqnarray}
\Delta v_b =-2\pi \rho_b
\,.
\end{eqnarray}
Assuming that the system occupies an area $\mathcal{A}_R$, the
background density can be written as
$\rho_b=-qN_b/\mathcal{A}_R=-qn_b$, where we have defined here
$n_b=N_b/\mathcal{A}_R$ the number density associated to the
background. For a neutral system $N_b=N$. The Coulomb potential of the
background can be obtained by solving Poisson equation with the
appropriate boundary conditions for each case. Also, it can be
obtained from the Green function computed in the previous section
\begin{equation}
  v_b(x,\varphi)=\int G(x,\varphi;x',\varphi') \rho_b \,dS'
  \,.
\end{equation}
This integral can be performed easily by using the Fourier series
decomposition~(\ref{eq:Fourier}) of the Green function $G$. Recalling
that $dS=\frac{1}{4}M^2 x (1+x^{-1})^4\,dx\,d\varphi$, after the angular
integration is done, only the zeroth order term in the Fourier series
survives
\begin{equation}
  v_b(x,\varphi)=\frac{\pi  \rho_b M^2}{2}
  \int_{1}^{x_m} g_0(x,x') \, x \left(1+\frac{1}{x}\right)^4 \,dx
  \,.
\end{equation}
The previous expression is for the half surface case and the grounded
horizon case. For the whole surface case, the lower limit of
integration should be replaced by $1/x_m$, or, equivalently, the
integral multiplied by a factor 2.

Using the explicit expressions for $g_0$,~(\ref{eq:g0-ws}),
(\ref{eq:g0-hs}), and (\ref{eq:g0-gh}) for each case, we find, for the
whole surface,
\begin{equation}
  v_b^{\mathrm{ws}}(x,\varphi)=-\frac{\pi \rho_b M^2}{8}
  \left[ h(x)-h(x_m) +2 p(x_m) \ln x_m - 4b_0 p(x_m)\right]
\end{equation}
where $p(x)$ was defined in equation~(\ref{eq:p(x)}), and
\begin{equation}
  h(x)=x^2+16x+\frac{16}{x}+\frac{1}{x^2}+12(\ln x)^2 - 34
  \,.
\end{equation}
Notice the following properties satisfied by the functions $p$ and $h$
\begin{equation}
  \label{eq:p-h-symmetry}
  p(x)=-p(1/x) \,,\qquad h(x)=h(1/x)
\end{equation}
and
\begin{equation}
  \label{eq:p-h-deriv}
  p(x)=x h'(x)/2 \,,\qquad p'(x)=2x\left(1+\frac{1}{x}\right)^4
\end{equation}
where the prime stands for the derivative.

The background potential for the half surface case, with the pair
potential $-\ln(|z-z'|M/2L)$ is
\begin{equation}
  v_b^{\mathrm{hs}}(x,\varphi)=-\frac{\pi \rho_b M^2}{8}
  \left[ h(x)-h(x_m) + 2 p(x_m) \ln\frac{x_m M}{2L} \right]
  \,.
\end{equation}
Also, the background potential in the half surface case, but with the pair
potential $-\ln(|z-z'|/\sqrt{|zz'|})+b_0$ is
\begin{equation}
  v_b^{\overline{\mathrm{hs}}}(x,\varphi)=
  -\frac{\pi \rho_b M^2}{8}
  \left[ h(x)-\frac{h(x_m)}{2}+p(x_m)\left(\ln\frac{x_m}{x}-2
    b_0\right)
    \right]\,.
\end{equation}

Finally, for the grounded horizon case,
\begin{equation}
  v_b^{\mathrm{gh}}(x,\varphi)=-\frac{\pi \rho_b M^2}{8}
  \left[ h(x) - 2 p(x_m) \ln x\right]
  \,.
\end{equation}

\section{Partition function and density at $\Gamma=2$}
\label{sec:correlations}

We will now show how at the special value of the coupling constant
$\Gamma=\beta q^2=2$ the partition function and $n$-body correlation
functions can be calculated exactly. 

In the following we will distinguish four cases labeled by $A$:
$A=\rm{hs}$, the plasma on the half surface (choosing $G^{\rm{hs}}$ as
the pair Coulomb potential); $A=\rm{ws}$, the plasma on the whole
surface (choosing $G^{\rm{ws}}$ as the pair Coulomb potential);
$A=\overline{\rm{hs}}$, the plasma on the half surface but with the
Coulomb potential $G^{\rm{ws}}$ of the whole surface case; and
$A=\rm{gh}$, the plasma on the half surface with the grounded horizon
(choosing $G^{\rm{gh}}$ as the pair Coulomb potential).
 
The total potential energy of the plasma is, in each case
\begin{eqnarray}
  \label{eq:hamiltonian-gen}
V^A=v_{0}^A+q\sum_i
v_{b}^A(x_i)+q^2\sum_{i<j}G^A(x_i,\varphi_i;x_j,\varphi_j)~, 
\end{eqnarray}
where $(x_i,\varphi_i)$ is the position of charge $i$ on the
surface, and
\begin{equation}
v_{0}^A=\frac{1}{2}\int \rho_b v_b^{A}(x,\varphi)\,dS
\end{equation}
is the self energy of the background in each of the four mentioned
cases. In the grounded case $A=\text{gh}$, one should add to
$V^{\text{gh}}$ in~(\ref{eq:hamiltonian-gen}) the self energy that
each particle has due to the polarization it creates on the grounded
conductor.

\subsection{The 2dOCP on half surface with potential $-\ln|z-z'|-\ln M/(2L)$}
\label{sec:half-surface-1}

\subsubsection{Partition function}

For this case, we work in the canonical ensemble with $N$ particles
and the background neutralizes the charges: $N_b=N$, and
$n=N/\mathcal{A}_R=n_b$. The potential energy of the system takes the
explicit form
\begin{eqnarray}
  V^{\mathrm{hs}} & = &
  -q^2\sum_{1\leq i<j\leq N}\ln|z_i-z_j|
  +\frac{q^2}{2}\alpha \sum_{i=1}^N h(x_i)
  +\frac{q^2}{2} N \ln \frac{M}{2L}
  -\frac{q^2}{4}N\alpha h(x_m)
  \nonumber\\
  &&+\frac{q^2}{2} N^2 \ln x_m
  -\frac{q^2}{4} \alpha^2 \int_{1}^{x_m} h(x) p'(x)\,dx 
  \label{eq:pot1}
\end{eqnarray}
where we have used the fact that $dS=\pi M^2 x (1+x^{-1})^4\,dx/2=\pi
M^2 p'(x)\,dx/4$, and we have defined
\begin{equation}
  \alpha=\frac{\pi n_b M^2}{4}\,.
\end{equation}
Integrating by parts the last term of~(\ref{eq:pot1}) and
using~(\ref{eq:p-h-deriv}), we find
\begin{eqnarray}
  V^{\mathrm{hs}} & = &
  -q^2\sum_{1\leq i<j\leq N}\ln|z_i-z_j|
  +\frac{q^2}{2}\alpha \sum_{i=1}^N h(x_i)
  +\frac{q^2}{2} N \ln \frac{M}{2L}
  +\frac{q^2}{2} N^2 \ln x_m
  \nonumber\\
  &&
  +\frac{q^2}{2}\alpha^2\int_{1}^{x_m} \frac{[p(x)]^2}{x}\,dx
  -\frac{q^2}{2} N \alpha h(x_m)
  \,.
  \label{eq:Vhs}
\end{eqnarray}

When $\beta q^2=2$, the canonical partition function can be written as
\begin{equation}
  Z^{\mathrm{hs}}=\frac{1}{\lambda^{2N}}\,Z_0^{\mathrm{hs}} 
  \exp(-\beta F_0^{\mathrm{hs}})
\end{equation}
with 
\begin{equation}
  \label{eq:F0}
  -\beta F_0^{\mathrm{hs}}=
   -N \ln \frac{M}{2L}
  - N^2 \ln x_m
  -\alpha^2\int_{1}^{x_m} \frac{[p(x)]^2}{x}\,dx
  + N \alpha h(x_m)
\end{equation}
and
\begin{equation}
  Z_0^{\mathrm{hs}}=\frac{1}{N!}\int
  \prod_{i=1}^N dS_{i}\, e^{-\alpha h(x_i)} 
  \prod_{1\leq i<j \leq N} |z_i-z_j|^2
  \,.
\end{equation}
where $\lambda=\sqrt{2\pi\beta\hbar^2/m}$ is the de Broglie thermal
wavelength. $Z_0$ can be computed using the original method for the
OCP in flat space~\cite{Jancovici81b,Alastuey81}, which was
originally introduced in the context of random matrices~\cite{Mehta91,
  Ginibre65}. By expanding the Vandermonde determinant
$\prod_{i<j}(z_i-z_j)$ and performing the integration over the angles,
the partition function can be written as
\begin{eqnarray} \label{cpf}
Z_0^{\mathrm{hs}}&=&
\prod_{k=0}^{N-1}{\cal B}_N(k)~,
\end{eqnarray}
where
\begin{eqnarray} 
{\cal B}_N(k)&=&
\int x^{2k} e^{-\alpha h(x)}\,dS
\\
&=&
\frac{\alpha}{n_b}\int_{1}^{x_m}
x^{2k} e^{-\alpha h(x)} p'(x)\,dx
\,.
\label{gamma}
\end{eqnarray}

In the flat limit $M\to0$, we have $x\sim 2r/M$, with $r$ the radial
coordinate of the flat space $\mathbb{R}^2$, and $h(x)\sim p(x)\sim
x^2$. Then, $\mathcal{B}_N$ reduces to
\begin{equation}
  \mathcal{B}_N(k)\sim\frac{1}{n_b \alpha^k}\, \gamma(k+1,N)
\end{equation}
where $\gamma(k+1,N)=\int_0^{N} t^k e^{-t}\,dt$ is the incomplete
Gamma function. Replacing into~(\ref{cpf}), we recover the partition
function for the OCP in a flat disk of radius $R$~\cite{Alastuey81}
\begin{equation}
  \ln Z^{\mathrm{hs}}=\frac{N}{2}\ln\frac{\pi L^2}{n_b\lambda^4}
  +\frac{3N^2}{4}-\frac{N^2}{2}\ln N
  +\sum_{k=1}^{N} \ln \gamma(k,N)
  \,.
\end{equation}

\subsubsection{Thermodynamic limit $R\to\infty$, $x_m\to\infty$, and fixed $M$}

Let us consider the limit of a large system when
$x_m=(\sqrt{R}+\sqrt{R-2M})^2/(2M)\to\infty$, $N\to\infty$, constant
density $n$, and constant $M$. Therefore $\alpha$ is also kept
constant.  In appendix~\ref{app:gamma}, we develop a uniform
asymptotic expansion of $\mathcal{B}_N(k)$ when $N\to\infty$ and
$k\to\infty$ with $(N-k)/\sqrt{N} = O(1)$. Let us define
$\hx_k$ by
\begin{equation}
  \label{eq:def-x_k}
  k=\alpha p(\hx_k)\,.
\end{equation}
The asymptotic expansion~(\ref{eq:asymptics-B}) of $\mathcal{B}_N(k)$
can be rewritten as
\begin{eqnarray}
  \label{eq:asympt-B}
  \mathcal{B}_N(k)&=&
  \frac{1}{2n_b}
  \sqrt{\pi \alpha \hx_k p'(\hx_k) }\,
  e^{2k\ln \hx_k -\alpha h(\hx_k)}\left[1+\erf\left(
    \epsilon_k\right)\right]
  \nonumber\\
  &&\times
  \left[1+\frac{1}{12 k}
    +\frac{1}{\sqrt{k}}\,\xi_1(\epsilon_k)
    +\frac{1}{k}\,\xi_2(\epsilon_k)\right]
\end{eqnarray}
where 
\begin{equation}
  \epsilon_k=\frac{2p(x_k)}{x_k p'(x_k)}\frac{N-k}{\sqrt{2k}}
\end{equation}
is a order one parameter, and the functions $\xi_1(\epsilon_k)$ and
$\xi_2(\epsilon_k)$ can be obtained from the calculation presented in
appendix~\ref{app:gamma}. They are integrable functions for
$\epsilon_k\in[0,\infty[$. We will obtain an expansion of the free
    energy up to the order $\ln N$. At this order the functions $\xi_{1,2}$
    do not contribute to the result.

Writing down
\begin{equation}
  \ln Z_0^{\mathrm{hs}}
  =\sum_{k=0}^{N} \ln \mathcal{B}_{N}(k)-\ln\mathcal{B}_{N}(N)
\end{equation}
and using the asymptotic expansion~(\ref{eq:asympt-B}), we have
\begin{eqnarray}
  \ln Z_0^{\mathrm{hs}}
  &=& -N\ln\frac{n_{b}}{\sqrt{2\pi}}+S_1^{\mathrm{hs}}
  +S_2^{\mathrm{hs}}
  +S_3^{\mathrm{hs}}
  +\frac{1}{12}\ln N
  \nonumber\\
  &&
  -\ln\left[\sqrt{\alpha} x_m \left(1+\frac{1}{x_m}\right)^2\right]
    -2N\ln x_m
    +\alpha h(x_m)
    +O(1)
\end{eqnarray}
with
\begin{eqnarray}
  S_1^{\mathrm{hs}}
  & = & \sum_{k=0}^N \ln \left( \sqrt{\alpha} 
   \hx_k \left(1+\frac{1}{\hx_k}\right)^2\right]
  \\
  S_2^{\mathrm{hs}}
  &=& \sum_{k=0}^{N} \left [2k\ln \hx_k - \alpha h(\hx_k) \right]
  \\
  S_3^{\mathrm{hs}}
  &=& \sum_{k=0}^N \ln
  \frac{1+\erf\left(\epsilon_k
    \right)}{2}
  \,.
\end{eqnarray}
Notice that the contribution of $\xi_{1}(\epsilon_k)$ is of order
one, since $\sum_k \xi_1(\epsilon_k)/\sqrt{k}\sim \int_0^{\infty}
\xi_{1}(\epsilon)\, d\epsilon = O(1)$. Also, $\sum_k
\xi_2(\epsilon_k)/k\sim (1/\sqrt{N})\int_0^{\infty}
\xi_{2}(\epsilon)\, d\epsilon = O(1/\sqrt{N})$.

$S_3^{\mathrm{hs}}$ gives a contribution of order $\sqrt{N}$,
transforming the sum over $k$ into an integral over the variable
$t=\epsilon_k$, we have
\begin{equation}
  S_3=\sqrt{2N} \int_0^{\infty} \ln
  \frac{1+\erf(t)}{2}\,dt + O(1) \,.
\end{equation}
This contribution is the same as the perimeter contribution in the
flat case.

To expand $S_1^{\mathrm{hs}}$ and $S_2^{\mathrm{hs}}$ up to order
$O(1)$, we need to use the Euler-McLaurin summation
formula~\cite{Abramowitz,Wong89}
\begin{equation}
  \sum_{k=0}^{N} f(k) = \int_0^N f(y)\,dy
  +\frac{1}{2}\left[f(0)+f(N)\right]
  +\frac{1}{12}\left[f'(N)-f'(0)\right]+\cdots
  \,.
\end{equation}
We find
\begin{eqnarray}
  S_1^{\mathrm{hs}}&=&
  \frac{N}{2}\ln\alpha + \alpha x_m^2 \left(\ln
  x_m-\frac{1}{2}\right)
  +\alpha x_m \left(8\ln x_m - 4\right)
\nonumber\\
&&
  +\left(14\alpha +\frac{1}{2} \right)\ln x_m
  + 6 \left(\ln x_m \right)^2
\end{eqnarray}
and
\begin{eqnarray}
  S_2^{\mathrm{hs}} &=&
  N^2 \ln x_m + N\ln x_m -\alpha N h(x_m)
  +\alpha^2 \int_{1}^{x_m} \frac{\left[p(x)\right]^2}{x}\,dx
  -\frac{\alpha}{2} \, h(x_m)
  +\frac{1}{6} \ln x_m
  \,.
\end{eqnarray}
Summing all terms in $\ln Z_0^{\mathrm{hs}}$ and those from $\beta
F_0^{\mathrm{hs}}$, we notice that all nonextensive terms cancel, as
it should be, and we obtain
\begin{equation}
  \ln Z^{\mathrm{hs}}=
  -N\beta f_B + 4 x_m \alpha - \mathcal{C}_R\,\beta\gamma_{\text{hard}}
  +\left(14\alpha - \frac{1}{6}\right) \ln x_m
  +O(1)
  \label{eq:Fhs-xm-infinity}
\end{equation}
where 
\begin{equation}
  \label{eq:bulk}
  \beta f_B=-\frac{1}{2}
  \ln\frac{2\pi^2 L^2}{n \lambda^4}
\end{equation}
is the bulk free energy of the OCP in the flat
geometry~\cite{Alastuey81}, 
\begin{eqnarray} \label{app:gam}
\beta\gamma_{\text{hard}}=-\sqrt{\frac{n_b}{2\pi}}
\int_0^\infty\ln\frac{1+\erf(y)}{2}\,dy
\end{eqnarray}
is the perimeter contribution to the free energy (``surface'' tension)
in the flat geometry near a plane hard wall~\cite{Jancovici94}, and
\begin{equation}
  \mathcal{C}_R=2\pi R=\pi M\sqrt{x_m p'(x_m)/2} = \pi M x_m
     + O(1)
\end{equation}
is the perimeter of the boundary at $x=x_m$.

The region $x\to\infty$ has zero curvature, therefore in the limit
$x_m\to\infty$, most of the system occupies an almost flat region. For
this reason, the extensive term (proportional to $N$) is expected to
be the same as the one in flat space $f_B$. The largest boundary of
the system $x=x_m$ is also in an almost flat region, therefore it is
not surprising to see the factor $\gamma_{\text{hard}}$ from the flat
geometry appear there as well. Nevertheless, we notice an additional
contribution $4\alpha x_m$ to the perimeter contribution, which comes
from the curvature of the system. In the logarithmic correction $\ln
x_m$, we notice a $-(1/6)\ln x_m$ term, the same as in a flat disk
geometry~\cite{Jancovici94}, but also a nonuniversal contribution due
to the curvature $14\alpha \ln x_m$.

\subsubsection{Thermodynamic limit at fixed shape: $\alpha\to\infty$ and
   $x_m$ fixed}

In the previous section we studied a thermodynamic limit case where a
large part of the space occupied by the particles becomes flat as
$x\to \infty$ keeping $M$ fixed. Another interesting thermodynamic
limit that can be studied is the one where we keep the shape of the
space occupied by the particles fixed. This limit corresponds to the
situation $M\to\infty$ and $R\to\infty$ while keeping the ratio $R/M$
fixed, and of course the number of particles $N\to\infty$ with the
density $n$ fixed. Equivalently, recalling that $N=\alpha p(x_m)$, in
this limit $x_m$ is fixed and finite, and $\alpha=\pi M^2
n_b/4\to\infty$. We shall use $\alpha$ as the large parameter for the
expansion of the free energy. In this limit, we expect the curvature
effects to remain important, in particular the bulk free energy
(proportional to $\alpha$) will not be the same as in flat space.

Using the expansion~(\ref{eq:BN-fixed-shape}) of $\mathcal{B}_N(k)$
for the fixed shape situation, we have
\begin{equation}
  \ln Z_0^{\mathrm{hs}}= N \ln\frac{\sqrt{\pi}}{n_b}+N
  \ln\sqrt{\alpha} +S_1^{\mathrm{hs,fixed}}+S_2^{\mathrm{hs,fixed}}
  +S_3^{\mathrm{hs,fixed}}+O(1)
\end{equation}
where now
\begin{eqnarray}
  \label{eq:S1-hs-fixed}
  S_1^{\mathrm{hs,fixed}}
  &=& \frac{1}{2}\sum_{k=0}^{N-1} \ln [\hx_k p'(\hx_k)] \\
  \label{eq:S2-hs-fixed}
  S_2^{\mathrm{hs,fixed}}
  &=& -\alpha\sum_{k=0}^{N-1} [h(\hx_k)-2p(\hx_k)\ln \hx_k]\\
  S_3^{\mathrm{hs,fixed}}
  &=& \sum_{k=0}^{N-1}\ln \frac{\erf(\epsilon_{k,1})+
    \erf(\epsilon_{k,m})}{2}
\end{eqnarray}
with $\epsilon_{k,m}$ and $\epsilon_{k,1}$ given in
equations~(\ref{eq:epsilon_km}) and~(\ref{eq:epsilon_k1}), and $\hx_k$
is given by $k=\alpha p(\hx_k)$. Using the Euler-McLaurin expansion, we
obtain
\begin{eqnarray}
  \label{eq:S1-hs-fixed-asympt}
  S_1^{\mathrm{hs,fixed}} &=& \alpha \int_1^{x_m}
  \frac{(1+x)^4}{x^3}\,\ln\frac{2(x+1)^4}{x^2}\,dx
  +O(1)\\
  \label{eq:S2-hs-fixed-asympt}
  S_2^{\mathrm{hs,fixed}}&=&N^2 \ln x_m -\alpha N h(x_m) 
  +\alpha^2 \int_{1}^{x_m} \frac{[p(x)]^2}{x}\,dx
  +\frac{\alpha}{2} h(x_m) - N\ln x_m + O(1)
  \,.
\end{eqnarray}
For $S_3^{\mathrm{hs,fixed}}$, the relevant contributions are
obtained when $k$ is of order $\sqrt{N}$, where $\epsilon_{k,1}$ is of
order one, and when $N-k$ is of order $\sqrt{N}$, where $\epsilon_{k,m}$
is of order one. In those regions, the sum can be changed into an integral
over the variable $t=\epsilon_{k,1}$ or $t=\epsilon_{k,m}$. This gives
\begin{eqnarray}
  S_3^{\mathrm{hs,fixed}}&=&-\sqrt{\frac{4\pi\alpha}{n_b}}\, 
  \left[x_m \left(1+\frac{1}{x_m}\right)^2
    +4\right]
    \beta\gamma_{\text{hard}} +O(1)
\end{eqnarray}
with $\gamma_{\text{hard}}$ given in equation~(\ref{app:gam}). Once again the
nonextensive terms (proportional to $\alpha^2$) in
$S_2^{\mathrm{hs,fixed}}$ cancel out with similar terms in
$F_0^{\mathrm{hs,fixed}}$ from equation~(\ref{eq:F0}). The final
result for the free energy $\beta F^{\mathrm{hs}}= -\ln
Z^{\mathrm{hs}}$ is
\begin{eqnarray}
  \ln Z^{\mathrm{hs}} &=&
  \alpha
  \left[ -p(x_m) \beta f_B 
    + \frac{1}{2} \left[ h(x_m) - 2 p(x_m) \ln x_m \right]
    +\int_{1}^{x_m} \frac{(1+x)^4}{x^3}\,\ln\frac{(x+1)^4}{x^2}\,dx
    \right]
  \nonumber\\
  &&
  -\sqrt{\frac{4\pi\alpha}{n_b}}\,
  \left[x_m \left(1+\frac{1}{x_m}\right)^2+4\right]
  \beta\gamma_{\text{hard}}
  +O(1)
  \label{eq:free-energy-fixed-shape}
\end{eqnarray}
where $f_B$, given by~(\ref{eq:bulk}), is the bulk free energy per
particle in a flat space. We notice the additional contribution to the
bulk free energy due to the important curvature effects [second and third
term of the first line of equation~(\ref{eq:free-energy-fixed-shape})]
that remain present in this thermodynamic limit.

The boundary terms, proportional to $\sqrt{\alpha}$, turn out to be
very similar to those of a flat space near a hard
wall~\cite{Jancovici82}, with a contribution $\beta
\gamma_{\text{hard}} \mathcal{C}_b$ for each boundary at $x_{b}=x_m$ and at
$x_{b}=1$ with perimeter 
\begin{equation}
  \label{eq:perimeter}
\mathcal{C}_b=\pi M \sqrt{\frac{x_b p'(x_b)}{2}}
=\pi M x_b \left(1+\frac{1}{x_b}\right)^2
\,.  
\end{equation}

Also, we notice the absence of $\ln \alpha$ corrections in the free
energy. This is in agreement with the general results from
Refs.~\cite{Jancovici94, Jancovici96}, where, using arguments from
conformal field theory, it is argued that for two-dimensional Coulomb
systems living in a surface of Euler characteristic $\chi$, in the
limit of a large surface keeping its shape fixed, the free energy
should exhibit a logarithmic correction $(\chi/6) \ln R$ where $R$ is
a characteristic length of the size of the surface. For our curved
surface studied in this section, the Euler characteristic is $\chi=0$,
therefore no logarithmic correction is expected.

\subsubsection{Distribution functions}

Following~\cite{Jancovici81b}, we can also find the $k$-body distribution
functions
\begin{eqnarray} \label{cf}
n^{(k){\mathrm{hs}}}(\qq_1,\ldots,\qq_k)= 
\det[{\cal K}_N^{\mathrm{hs}}(\qq_i,\qq_j)]_{(i,j)\in\{1,\ldots,k\}^2}~,
\end{eqnarray}
where $\qq_i=(x_i,\varphi_i)$ is the position of the particle $i$, and
\begin{eqnarray} \label{KN}
{\cal K}_N^{\mathrm{hs}}(\qq_i,\qq_j)
=\sum_{k=0}^{N-1}\frac{z_{i}^{k}\bar{z}_j^{k} 
e^{-\alpha[h(|z_i|)+h(|z_j|)]/2}}{{\cal B}_N(k)}~.
\end{eqnarray}
where $z_k=x_k e^{i\varphi_k}$. In particular, the one-body density is
given by
\begin{equation}
  \label{eq:density-Nfinite}
  n^{\mathrm{hs}}(x)=\mathcal{K}_N(\qq,\qq)=
  \sum_{k=0}^{N-1} \frac{x^{2k} e^{-\alpha
      h(x)}}{\mathcal{B}_N(k)}
\,.
\end{equation}

\subsubsection{Internal screening}
Internal screening means that at equilibrium, a particle of the system
is surrounded by a polarization cloud of opposite charge. It is usually
expressed in terms of the simplest of the multipolar sum rules
\cite{Martin88}: the charge or electroneutrality sum rule, which for
the OCP reduces to the relation
\begin{eqnarray} \label{csr}
\int n^{(2){\mathrm{hs}}}(\qq_1,\qq_2)\,dS_2=(N-1)n^{(1){\mathrm{hs}}}(\qq_1)~,
\end{eqnarray}
This relation is trivially satisfied because of the particular
structure~(\ref{cf}) of the correlation function expressed as a
determinant of the kernel $\mathcal{K}_{N}^{\mathrm{hs}}$, and the
fact that $\mathcal{K}_{N}^{\mathrm{hs}}$ is a projector
\begin{equation}
  \int dS_3\, \mathcal{K}_N^{\mathrm{hs}}(\qq_1,\qq_3) 
  \mathcal{K}_N^{\mathrm{hs}}(\qq_3,\qq_2)
  = \mathcal{K}_N^{\mathrm{hs}}(\qq_1,\qq_2)
  \,.
\end{equation}
Indeed,
\begin{eqnarray} \nonumber
\int n^{(2){\mathrm{hs}}}(\qq_1,\qq_2)\,dS_2&=&
\int
[{\cal K}_N^{\mathrm{hs}}(\qq_1,\qq_1)
  {\cal K}_N^{\mathrm{hs}}(\qq_2,\qq_2)-
{\cal K}_N^{\mathrm{hs}}(\qq_1,\qq_2)
{\cal K}_N^{\mathrm{hs}}(\qq_2,\qq_1)]\,dS_2
\\ \nonumber
&=&\int n^{(1){\mathrm{hs}}}(\qq_1)n^{(1){\mathrm{hs}}}(\qq_2)\,dS_2-
{\cal K}_N^{\mathrm{hs}}(\qq_1,\qq_1)
\\ 
&=&
(N-1)n^{(1){\mathrm{hs}}}(\qq_1)
\,.
\end{eqnarray}

\subsubsection{External screening}

External screening means that, at equilibrium, an external charge
introduced into the system is surrounded by a polarization cloud of
opposite charge.  When an external infinitesimal point charge $Q$ is
added to the system, it induces a charge density
$\rho_Q(\qq)$. External screening means that
\begin{eqnarray} \label{es}
  \int \rho_Q(\qq)\, dS=-Q~.
\end{eqnarray}
Using linear response theory we can calculate $\rho_Q$ to first order
in $Q$ as follows. Imagine that the charge $Q$ is at $\qq$. Its
interaction energy with the system is $\hat{H}_{int}=Q\hat{\phi}(\qq)$
where $\hat{\phi}(\qq)$ is the microscopic electric potential created
at $\qq$ by the system. Then, the induced charge density at $\qq'$ is
\begin{eqnarray}
 \rho_Q(\qq')=-\beta\langle\hat{\rho}(\qq')\hat{H}_{int}\rangle_T=
 -\beta Q \langle\hat{\rho}(\qq')\hat{\phi}(\qq)\rangle_T~,
\end{eqnarray}
where $\hat{\rho}(\qq')$ is the microscopic charge density at $\qq'$,
$\langle AB\rangle_T=\langle AB\rangle-\langle A\rangle\langle
B\rangle$, and $\langle\ldots\rangle$ is the thermal average.
Assuming external screening~(\ref{es}) is satisfied, one obtains the
Carnie-Chan sum rule \cite{Martin88}
\begin{eqnarray} \label{cc}
  \beta\int \langle\hat{\rho}(\qq')\hat{\phi}(\qq)\rangle_T\,dS'=1~.
\end{eqnarray}
Now in a uniform system starting from this sum rule one can derive the
second moment Stillinger-Lovett sum rule \cite{Martin88}. This is not
possible here because our system is not homogeneous since the
curvature is not constant throughout the surface but varies from point
to point.  If we apply the Laplacian respect to $\qq$ to this
expression and use Poisson equation
\begin{eqnarray}
  \Delta_{\qq}\langle\hat{\rho}(\qq')\hat{\phi}(\qq)\rangle_T=
  -2\pi\langle\hat{\rho}(\qq')\hat{\rho}(\qq)\rangle_T~,
\end{eqnarray}
we find
\begin{eqnarray} \label{csr'}
  \int \rho_e^{(2)}(\qq',\qq)\,dS'=0~,
\end{eqnarray}
where
$\rho_e^{(2)}(\qq',\qq)=\langle\hat{\rho}(\qq')\hat{\rho}(\qq)\rangle_T$
is the excess pair charge density function. Eq.~(\ref{csr'}) is
another way of writing the charge sum rule Eq.~(\ref{csr}) in the
thermodynamic limit.


\subsubsection{Asymptotics of the density in the
  limit $x_m\to\infty$ and $\alpha$ fixed, for $1\ll x \ll x_m$}

The formula (\ref{eq:density-Nfinite}) for the one-body density,
although exact, does not allow a simple evaluation of the density at a
given point in space, as one has first to calculate $\mathcal{B}_N(k)$
through an integral and then perform the sum over $k$. One can then
try to determine the asymptotic behaviors of the density.

In this section, we consider the limit $x_m\to\infty$ and $\alpha$
fixed, and we study the density in the bulk of the system $1\ll x\ll
x_m$.

In the sum~(\ref{eq:density-Nfinite}), the dominant terms are the ones
for which $k$ is such that $\hx_k=x$, with $\hx_k$ defined
in~(\ref{eq:def-x_k}). Since $1\ll x \ll x_m$, the dominant terms
in the calculation of the density are obtained for values of $k$ such
that $1\ll k \ll N$. Therefore in the limit $N\to\infty$, in the
expansion~(\ref{eq:asympt-B}) of $B_N(k)$, the argument of the error
function is very large, then the error function can be replaced by
1. Keeping the correction $1/(12 k)$ from~(\ref{eq:asympt-B}) allow us
to obtain an expansion of the density up to terms of order
$O(1/x^2)$. Replacing the sum over $k$ into an integral over $\hx_k$, we
have
\begin{equation}
  n^{\text{hs}}(x)=\frac{n_b}{\sqrt{\pi}}
  \int_{-\infty}^{\infty}
  e^{\Psi(\hx_k)} f(\hx_k) \left(1-\frac{1}{12\alpha p(\hx_k)} \right)
  \,d\hx_k
\end{equation}
with
\begin{equation}
  \Psi(\hx_k)=2\alpha p(\hx_k)\ln \frac{x}{\hx_k}
  -\alpha[h(x)-h(\hx_k)]
\end{equation}
and
\begin{equation}
  f(\hx_k)=\sqrt{\frac{\alpha p'(\hx_k)}{\hx_k}}
  \,.
\end{equation}
We proceed now to use the Laplace method to compute this integral. The
function $\Psi(\hx_k)$ has a maximum for $x=\hx_k$, with $\Psi(x)=0$ and
\begin{subequations}
  \label{eq:Psis}
\begin{eqnarray}
\Psi''(x)&=&-\frac{2\alpha p'(x)}{x}\\
\Psi^{(3)}(x)&=&-\frac{4\alpha}{x}+O(1/x^2)\\
\Psi^{(4)}(x)&=&\frac{4\alpha}{x^2}+O(1/x^3)
\,.
\end{eqnarray}  
\end{subequations}
Expanding for $\hx_k$ close to $x$ and for $x\gg1$ up to order $1/x^2$,
we have
\begin{eqnarray}
   n^{\text{hs}}(x)&=&\frac{n_b}{\sqrt{\pi}}
   \int_{-\infty}^{+\infty}
   e^{-\alpha p'(x) (\hx_k-x)^2/x}
   \left( f(x)+f'(x)(\hx_k-x)+\frac{f''(x)(\hx_k-x)^2}{2}
   \right)
   \nonumber\\
   &&\times\left(1+\frac{1}{3!} \Psi^{(3)}(x)(\hx_k-x)^3
   +\frac{1}{4!}\Psi^{(4)}(x)(\hx_k-x)^4
   +\frac{[\Psi^{(3)}(x)]^2}{3!^2\ 2}(\hx_k-x)^6
   \right)
   \nonumber\\
   &&
   \times
   \left(1-\frac{1}{12\alpha p(x)}+O(1/x^3)\right)
   \,d\hx_k
   \,.
\end{eqnarray}
For the expansion of $f(\hx_k)$ around $\hx_k=x$, it is interesting to
notice that 
\begin{equation}
  f'(x)=O(1/x^2)\,, \qquad\text{and }f''(x)=O(1/x^3)\,.
\end{equation}
In the integral, the factor containing $f'(x)$ is multiplied $(\hx_k-x)$
which after integration vanishes. Therefore, the relevant contributions
to order $O(1/x^2)$ are
\begin{eqnarray}
   n^{\text{hs}}(x)&=&\frac{n_b}{\sqrt{\pi}}
   \int_{-\infty}^{+\infty}
   e^{-\alpha p'(x) (\hx_k-x)^2/x} \sqrt{\frac{\alpha p'(x)}{x}}
   \nonumber\\
   &&\times\left(1+\frac{1}{3!} \Psi^{(3)}(x)(\hx_k-x)^3
   +\frac{1}{4!}\Psi^{(4)}(x)(\hx_k-x)^4
   +\frac{[\Psi^{(3)}(x)]^2}{3!^2\ 2}(\hx_k-x)^6
   \right)
   \nonumber\\
   &&
   \times
   \left(1-\frac{1}{12\alpha p(x)}\right)
   \,d\hx_k+O(1/x^3)
   \,.
\end{eqnarray}
Then, performing the Gaussian integrals and replacing the dominant
values of $\Psi(x)$ and its derivatives from Eqs.~(\ref{eq:Psis}) for
$x\gg1$, we find
\begin{equation}
  n(x)=n_b   \left(1+\frac{1}{12\alpha x^2}\right)
  \left(1-\frac{1}{12\alpha x^2}\right) +O(1/x^3)
  =n_b + O(1/x^3)
  \,.
\end{equation}
In the bulk of the plasma, the density of particles equal the bulk
density, as expected. The above calculation, based the Laplace method,
generates an expansion in powers of $1/x$ for the density. The first
correction to the background density, in $1/x^2$, has been shown to be
zero. We conjecture that this is probably true for any subsequent
corrections in powers $1/x$ if the expansion is pushed further,
because the corrections to the bulk density are probably exponentially
small, rather than in powers of $1/x$, due to the screening
effects. In the following subsections, we consider the expansion of
the density in other types of limits, and in particular close to the
boundaries, and the results suggest that our conjecture is true.


\subsubsection{Asymptotics of the density close to the boundary in the
  limit $x_m\to\infty$}

We study here the density close to the boundary $x=x_m$ in the limit
$x_m\to\infty$ and $M$ fixed. Since in this limit this region is
almost flat, one would expect to recover the result for the OCP in a
flat space near a wall~\cite{Jancovici82}. Let $x=x_m+y$ where
$y\ll x_m$ is of order 1.

Using the dominant term of the asymptotics~(\ref{eq:asympt-B}),
\begin{eqnarray}
  \label{eq:asympt-B-domin}
  \mathcal{B}_N(k)&=&
  \frac{1}{2n_b}
  \sqrt{\pi \alpha \hx_k p'(\hx_k) }\,
  e^{2k\ln \hx_k -\alpha h(\hx_k)}\left[1+\erf\left(
    {\epsilon_k}\right)\right]
  \,,
\end{eqnarray}
we have
\begin{equation}
  n^{\mathrm{hs}}(x)= \frac{2n_b}{\sqrt{\pi}}
    \sum_{k=0}^{N-1}
    \frac{e^{2k(\ln x-\ln \hx_k)-\alpha [h(x)-h(\hx_k)]}}{
      \sqrt{\alpha \hx_k p' (\hx_k)} \left[
        1+\erf\left({\epsilon_k}\right)
        \right]}
\end{equation}
where we recall that $\hx_k=p^{-1}(k/\alpha)$. The exponential term in
the sum has a maximum when $\hx_k=x$ {\sl i.e.\/} $k=k_{\max}=\alpha
p(x)$, and since $x$ is close to $x_m\to\infty$, the function is very
peaked near this maximum. Thus, we can use Laplace method to compute
the sum. Expanding the argument of the exponential up to order 2 in
$k-k_{\max}$, we have
\begin{equation}
  n^{\mathrm{hs}}(x)=\frac{2n_b}{\sqrt{\pi}}
    \sum_{k=0}^{N-1}
    \frac{\exp\left[-\frac{2}{\alpha x p'(x)} (k-k_{\max})^2
        \right]}{\sqrt{\alpha x p'(x)}\left[
        1+\erf\left(\epsilon_k\right)
        \right] }
\end{equation}
Now, replacing the sum by an integral over $t=\epsilon_k$ and
replacing $x=x_m-y$, we find
\begin{equation}
  \label{eq:density-close-border}
  n^{\mathrm{hs}}(x)=\frac{2n_b}{\sqrt{\pi}}
    \int_0^{\infty}
    \frac{\exp\left[-(t-\sqrt{2\alpha}y)^2\right]}{1+\erf(t)}
    \,dt
    \,.
\end{equation}
Since both $x_m\to\infty$, and $x\to\infty$, in that region, the space
is almost flat. If $s$ is the geodesic distance from $x$ to the
border, then we have $y\sim\sqrt{(\pi n_b/\alpha)}\, s$, and
equation~(\ref{eq:density-close-border}) reproduces the result for the
flat space~\cite{Jancovici82}, as expected.

\subsubsection{Density in the thermodynamic limit at fixed shape:
  $\alpha\to\infty$ and $x_m$ fixed.}
\label{sec:density-canonic-fixed-shape}

Using the expansion~(\ref{eq:BN-fixed-shape}) of $\mathcal{B}_N(k)$ for the
fixed shape situation, we have
\begin{equation}
  n^{\mathrm{hs}}(x)=2n_b
  \sum_{k=0}^{N-1}
  \frac{e^{-\alpha [h(x)-2p(\hx_k)\ln x- h(\hx_k) + 2p(\hx_k) \ln \hx_k]}
    }{
    \sqrt{\alpha \pi \hx_k p'(\hx_k)}
    \left[\erf(\epsilon_{k,1})+\erf(\epsilon_{k,m})\right]
    }
  \,.
  \label{eq:density-fixed-shape-start1}
\end{equation}
Once again, to evaluate this sum when $\alpha\to\infty$ it is
convenient to use Laplace method. The argument of the exponential has
a maximum when $k$ is such that $\hx_k=x$. Transforming the sum into
an integral over $\hx_k$, and expanding the argument of the integral
to order $(\hx_k-x)^2$, we have
\begin{equation}
  n^{\mathrm{hs}}(x)=\frac{2n_b\sqrt\alpha}{\sqrt{\pi}}
  \int_{1}^{x_m}
  \sqrt{\frac{p'(\hx_k)}{\hx_k}}
  \frac{e^{-\alpha p'(x) (x-\hx_k)^2/x}
    }{
    \erf(\epsilon_{k,1})+\erf(\epsilon_{k,m})
    }
  \,d\hx_k
  \,.
  \label{eq:density-fixed-shape-start2}
\end{equation}

Depending on the value of $x$ the result will be different, since we
have to take special care of the different cases when the
corresponding dominant values of $\hx_k$ are close to the limits of 
integration or not.

Let us first consider the case when $x-1$ and $x_m-x$ are of order
one. This means we are interested in the density in the bulk of the
system, far away from the boundaries. In this case, since
$\epsilon_{k,1}$ and $\epsilon_{k,m}$, defined
in~(\ref{eq:epsilon_km}) and~(\ref{eq:epsilon_k1}), are proportional
to $\sqrt{\alpha}\to\infty$, then each error function in the
denominator of~(\ref{eq:density-fixed-shape-start2}) converge to
1. Also, the dominant values of $\hx_k$, close to $x$ (more precisely,
$x-\hx_k$ of order $1/\sqrt{\alpha}$), are far away from $1$ and $x_m$
(more precisely, $\hx_k-1$ and $x_m-\hx_k$ are of order 1). Then, we can
extend the limits of integration to~$-\infty$~and~$+\infty$, and
approximate $\hx_k$ by $x$ in the term $p'(\hx_k)/\hx_k$. The resulting
Gaussian integral is easily performed, to find
\begin{equation}
  \label{eq:density-bulk}
  n(x)=n_b\,,
  \qquad \text{when $x-1$ and $x_m-x$ are of order 1.}
\end{equation}

Let us now consider the case when $x-x_m$ is of order
$1/\sqrt{\alpha}$, {\sl i.e.\/} we study the density close to the
boundary at $x_m$. In this case $\epsilon_{k,m}$ is of order 1 and the
term $\erf(\epsilon_{k,m})$ cannot be approximated to 1, whereas
$\epsilon_{k,1}\propto\sqrt{\alpha}\to\infty$ and
$\erf(\epsilon_{k,1})\to 1$. The terms $p'(\hx_k)/\hx_k$ and $p'(x)/x$
can be approximated to $p'(x_m)/x_m$ up to corrections of order
$1/\sqrt{\alpha}$. Using $t=\epsilon_{k,m}$ as new variable of
integration, we obtain
\begin{equation}
  \label{eq:density-border-xm}
  n^{\mathrm{hs}}(x)=\frac{2 n_b}{\sqrt{\pi}}
  \int_0^{+\infty}
  \frac{\exp\left[-\left(t-\sqrt{\frac{\alpha p'(x_m)}{x_m}}(x_m-x)
        \right)^2\right]}{1+\erf(t)}
    \,dt\,,
    \quad
    \text{for $x_m-x$ of order }\frac{1}{\sqrt{\alpha}}
    \,.
\end{equation}
In the case where $x-1$ is of order $1/\sqrt{\alpha}$, close to the
other boundary, a similar calculation yields,
\begin{equation}
  \label{eq:density-border-1}
  n^{\mathrm{hs}}(x)=\frac{2 n_b}{\sqrt{\pi}}
  \int_0^{+\infty}
  \frac{\exp\left[-\left(t-\sqrt{\alpha p'(1)}(x-1)
        \right)^2\right]}{1+\erf(t)}
    \,dt\,,
    \quad
    \text{for $x-1$ of order }\frac{1}{\sqrt{\alpha}}
    \,.
\end{equation}
where $p'(1)=32$. 

\begin{figure}
\begin{center}
\includegraphics[width=\GraphicsWidth]{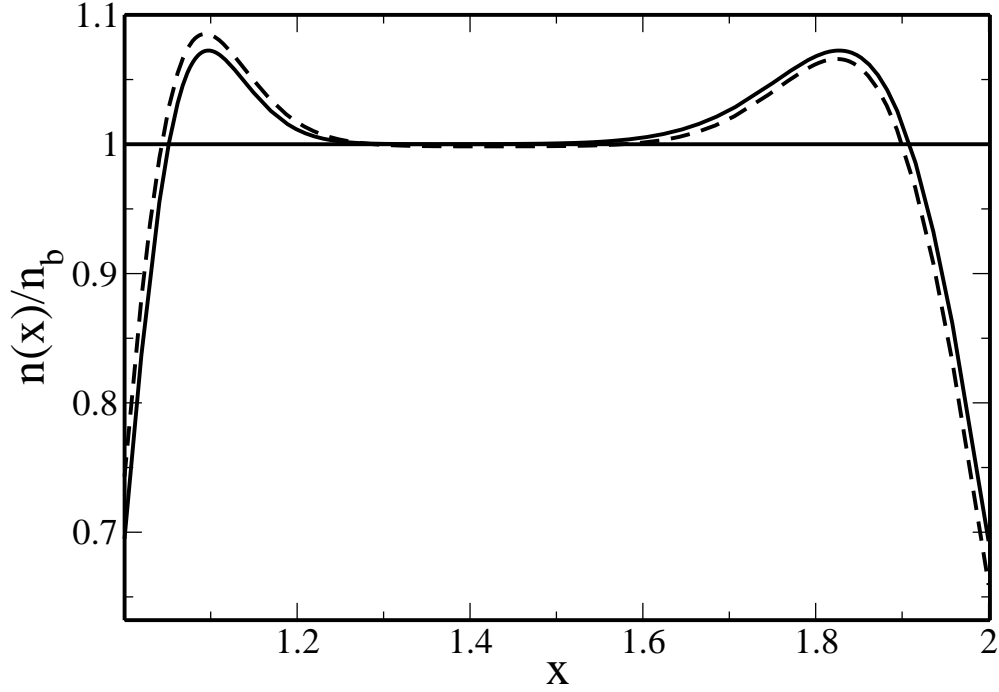}
\end{center}
\caption{The normalized one-body density $n^{\text{hs}}(x)/n_b$, for
  the 2dOCP on just one universe of the surface ${\cal S}$. The dashed
  line corresponds to a numerical evaluation, obtained
  from~(\ref{eq:density-Nfinite}), with $N=100$, $x_m=2$ and
  $\alpha=4.15493$. The full line corresponds to the asymptotic result
  in the fixed shape limit when $\alpha\to\infty$, and $x_m=2$
  fixed. }
\label{fig:density-fixed-shape-hs}
\end{figure}

Fig.~\ref{fig:density-fixed-shape-hs} compares the density profile for
finite $N=100$ with the asymptotic
results~(\ref{eq:density-bulk}), (\ref{eq:density-border-xm})
and~(\ref{eq:density-border-1}). The figure show how the density tends
to the background density, $n_{b}$, far from the boundaries. Near the
boundaries it has a peak, eventually decreasing below $n_{b}$ when
approaching the boundary. In the limit $\alpha\to\infty$, the value of
the density at each boundary is $n_b\ln 2$.

Interestingly, the results~(\ref{eq:density-bulk}),
(\ref{eq:density-border-xm}) and~(\ref{eq:density-border-1}) turn out
to be the same than the one for a flat space near a hard
wall~\cite{Jancovici82}.  From the metric~(\ref{eq:metric-in-x}), we
deduce that the geodesic distance to the boundary at $x_m$ is
$s=M(x_m-x)\sqrt{p'(x_m)/(8x_m)}$ (when $x_m-x$ is of order
$1/\sqrt{\alpha}$), and a similar expression for the distance to the
boundary at $x=1$ replacing $x_m$ by 1. Then, in terms of the geodesic
distance $s$ to the border, the results~(\ref{eq:density-border-xm})
and~(\ref{eq:density-border-1}) are exactly the same as those of an
OCP in a flat space close to a plane hard wall~\cite{Jancovici82},
\begin{equation}
  \label{eq:density-border-flat}
  n(s)=\frac{2 n_b}{\sqrt{\pi}}
  \int_0^{+\infty}
  \frac{\exp\left[-\left(t-s\sqrt{2\pi n_b}
        \right)^2\right]}{1+\erf(t)}
    \,dt\,.
\end{equation}

This result shows that there exists an interesting universality for
the density, because, although we are considering a limit where
curvature effects are important, the density turns out to be the same
as the one for a flat space.

\subsection{The 2dOCP on the whole surface with potential 
$-\ln(|z-z'|/\sqrt{|zz'|})$}

\subsubsection{Partition function}

Until now we studied the 2dOCP on just one universe. Let us find the
thermodynamic properties of the 2dOCP on the whole surface ${\cal
  S}$. In this case, we also work in the canonical ensemble with a
global neutral system. The position $z_k=x_k e^{i\varphi_k}$ of each
particle can be in the range $1/x_m<x_k<x_m$. The total number
particles $N$ is now expressed in terms of the function $p$ as
$N=2\alpha p(x_m)$. Similar calculations to the ones of the previous
section lead to the following expression for the partition function,
when $\beta q^2=2$,
\begin{equation}
  Z^{\mathrm{ws}}=\frac{1}{\lambda^{2N}}Z_0^{\mathrm{ws}}
  \exp(-\beta F_0^{\mathrm{ws}})
\end{equation}
now, with
\begin{equation}
  -\beta F_0^{\mathrm{ws}}
  = N b_0+N\alpha h(x_m)- \frac{N^2}{2} \ln x_m -\alpha^2
  \int_{1/x_m}^{x_m} \frac{\left[p(x)\right]^2}{x}\,dx
\end{equation}
and
\begin{equation}
  Z_0^{\mathrm{ws}}=\frac{1}{N!}\int
  \prod_{i=1}^N dS_{i}\, e^{-\alpha h(x_i)} x_{i}^{-N+1} 
  \prod_{1\leq i<j \leq N} |z_i-z_j|^2
  \,.
\end{equation}
Expanding the Vandermonde determinant and performing the angular
integrals we find
\begin{equation}
  Z_0^{\mathrm{ws}}=\prod_{k=0}^{N-1} 
  \tilde{\mathcal{B}}_N(k)
\end{equation}
with
\begin{eqnarray} 
\tilde{{\cal B}}_N(k)&=&
\int x^{2k-N+1} e^{-\alpha h(x)}\,dS
\\
&=&
\frac{\alpha}{n}\int_{1/x_m}^{x_m}
x^{2k-N+1} e^{-\alpha h(x)} p'(x)\,dx
\,.
\label{gamma-tilde}
\end{eqnarray}
The function $\tilde{\mathcal{B}}_N(k)$ is very similar to
$\mathcal{B}_{N}$, and its asymptotic behavior for large values of $N$
can be obtained by Laplace method as explained in appendix~\ref{app:gamma}.

\subsubsection{Thermodynamic limit $R\to\infty$, $x_m\to\infty$, and fixed $M$}

Writing the partition function as 
\begin{equation}
  \ln Z_0^{\mathrm{ws}} = \sum_{k=0}^{N} \ln \tilde{\mathcal{B}}_N(k)
  -\ln \tilde{\mathcal{B}}_N(N)
  \,,
\end{equation}
and using the asymptotic expansion~(\ref{eq:asympt-tildeBN}) for
$\tilde{\mathcal{B}}_N$, we have
\begin{eqnarray}
  \ln Z_0^{\mathrm{ws}}&=&
  -\ln\frac{n_b}{\sqrt{2\pi}}
  +S_1^{\mathrm{ws}}+S_2^{\mathrm{ws}}+S_3^{\mathrm{ws}}
  +S_4^{\mathrm{ws}}+S_5^{\mathrm{ws}}
  -\ln\left[\sqrt{\alpha}\, x_m \left(1+\frac{1}{x_m}\right)^2 \right]
  \nonumber\\
  &&-\ln x_m
  -N\ln x_m
  +\alpha h(x_m)
\end{eqnarray}
where
\begin{eqnarray}
  S_1^{\mathrm{ws}}
  &=&\sum_{k=0}^{N} \ln \left[\sqrt{\alpha}\,\hx_{k-\frac{N}{2}}
    \left(1+\frac{1}{\hx_{k-\frac{N}{2}}}\right)^2\right]\\
  S_2^{\mathrm{ws}}
  &=&\sum_{k=0}^{N} 2\left(k-\frac{N}{2}\right)\ln
  \hx_{k-\frac{N}{2}}
  -\alpha h(\hx_{k-\frac{N}{2}})\\
  S_3^{\mathrm{ws}}&=&\sum_{k=0}^N \ln
  \frac{\erf(\epsilon_{k,\min})+\erf(\epsilon_{k,\max})}{2}
  \\
  S_4^{\mathrm{ws}}&=&\sum_{k=0}^{N} \ln \hx_{k-\frac{N}{2}}
  \\
  S_5^{\mathrm{ws}}&=&\sum_{k'=1}^{N/2}
  \left(\frac{1}{12}+\frac{3}{8}\right)\frac{1}{|k'|}
  +
  \sum_{k'=-N/2}^{-1}
  \left(\frac{1}{12}-\frac{1}{8}\right)\frac{1}{|k'|}
  =\frac{5}{6}\ln x_m + O(1)
\end{eqnarray}
and $\epsilon_{k,\min}$ and $\epsilon_{k,\max}$ are defined in
equation~(\ref{eq:epsilons-min-max}). Notice that $S_4^{\text{ws}}=0$
due to the symmetry relation $\hx_{-\ell}=1/\hx_{\ell}$, therefore only
the sums $S_1^{\text{ws}}$, $S_2^{\text{ws}}$, $S_3^{\text{ws}}$ and
$S_5^{\text{ws}}$ contribute to the result. These sums are similar to
the ones defined for the half surface case, with the difference that
the running index $k'=k-N/2$ varies from $-N/2$ to $N/2$ instead of
$0$ to $N$ as in the half surface case. This difference is important
when considering the remainder terms in the Euler-McLaurin expansion,
because now both terms for $k'=-N/2$ and $k'=N/2$ are important in the
thermodynamic limit. In the half surface case only the
contribution for $k=N$ was important in the thermodynamic limit.

The asymptotic expansion of each sum, for $x_m\to\infty$, is now
\begin{eqnarray}
  S_1^{\mathrm{ws}}&=&\frac{N}{2}\ln \alpha + 
  x_m^2(2\ln x_m-1) +2x_m(8\ln x_m-4)
  +(28\alpha+1) \ln x_m+12\alpha(\ln x_m)^2
  +O(1)
  \nonumber\\
  \\
  S_2^{\mathrm{ws}}&=&\frac{N^2}{2}\ln x_m +\alpha^2 \int_{1/x_m}^{x_m}
  \frac{\left[p(x)\right]^2}{x}\,dx
  -\alpha N h(x_m)
  +N\ln x_m -\alpha h(x_m) +\frac{1}{3}\ln x_m +O(1)
  \nonumber\\
  \\
  S_3^{\mathrm{ws}}&=&-2 x_m \sqrt{\frac{4\pi\alpha}{n_b}}\,
  \beta\gamma_{\text{hard}} +O(1)
\end{eqnarray}
where $\gamma$ is defined in equation~(\ref{app:gam}). The free energy
is given by $\beta F^{\mathrm{ws}}=-\ln Z^{\mathrm{ws}}$, with
\begin{eqnarray}
  \ln Z^{\mathrm{ws}}&=&2\alpha x_m^2 \ln x_m+ N \left(b_0+
  \ln\frac{\sqrt{2\pi\alpha}}{\lambda^2 n_b}\right)
  -\alpha x_m^2 
  +8\alpha x_m (2 \ln x_m -1)
  -2\mathcal{C}_R \,\beta \gamma_{\text{hard}}
  \nonumber\\
  &&
  +12 \alpha (\ln x_m)^2+28 \alpha \ln x_m +\frac{1}{6}\ln x_m
  +O(1)\,.
\end{eqnarray}
We notice that the free energy for this system turns out to be
nonextensive with a term $2x_m^2 \ln x_m$. This is probably due to the
special form of the potential $-\ln(|z-z'|/\sqrt{|zz'|})$: the
contribution from the denominator in the logarithm can be written as a
one-body term $[(N-1)/2]\ln x $, which is not intensive but extensive.
However, this nonextensivity of the final result is mild, and can be
cured by choosing the arbitrary additive constant $b_0$ of the Coulomb
potential as $b_0=-\ln (M x_m)+\text{constant}$.

\subsubsection{Thermodynamic limit at fixed shape: 
$\alpha\to\infty$ and $x_m$ fixed}

For this situation, we use the asymptotic
behavior~(\ref{eq:tildeBN-fixed-shape}) of $\tilde{\mathcal{B}}_N$
\begin{equation}
  \ln Z_0^{\mathrm{ws}}= N \ln \frac{\sqrt{\pi \alpha}}{n_b}+ 
  S_1^{\mathrm{ws,fixed}}+S_2^{\mathrm{ws,fixed}}
  +S_3^{\mathrm{ws,fixed}}+S_4^{\mathrm{ws,fixed}}
\end{equation}
where, now
\begin{eqnarray}
  S_1^{\mathrm{ws,fixed}} &=& \frac{1}{2}\sum_{k=0}^{N-1} \ln [\hx_{k-\frac{N}{2}}
 p'(\hx_{k-\frac{N}{2}})] \\
  S_2^{\mathrm{ws,fixed}} &=& -\alpha\sum_{k=0}^{N-1}
  [h(\hx_{k-\frac{N}{2}})-2p(\hx_{k-\frac{N}{2}})
    \ln \hx_{k-\frac{N}{2}}]\\
  S_3^{\mathrm{ws,fixed}} &=& \sum_{k=0}^{N-1}\ln \frac{\erf(\epsilon_{k,\min})+
    \erf(\epsilon_{k,\max})}{2}\\
  S_4^{\mathrm{ws,fixed}}&=&\sum_{k=0}^{N-1} \ln \hx_{k-\frac{N}{2}}
\end{eqnarray}
These sums can be computed as earlier using Euler-McLaurin summation
formula. We notice that
\begin{equation}
  S_4^{\mathrm{ws,fixed}}
  =\alpha \int_{1/x_m}^{x_m} \ln x\, p'(x)\,dx + O(1) =0+ O(1)
\end{equation}
because of the symmetry properties $\ln(1/x)=-\ln x$ and
$p'(1/x)d(1/x)=-p'(x)dx$. In the computation of
$S_2^{\mathrm{ws,fixed}}$ there is an important difference with the
case of the half surface section, due to the contribution when $k=0$,
since $\hx_{-N/2}=1/\hx_{N/2}=1/x_m$
\begin{eqnarray}
  S_2^{\mathrm{ws,fixed}}=-\alpha N h(x_m) - \frac{N^2}{2}\ln x_m
  +\alpha^2 \int_{1/x_m}^{x_m} \frac{\left[p(x)\right]^2}{x}\,dx
  +O(1)
  \,.
\end{eqnarray}
There is no $O(\alpha)$ contribution from
$S_2^{\mathrm{ws,fixed}}$. Finally, the free energy $\beta
F^{\mathrm{ws}}=-\ln Z^{\mathrm{ws}}$ is given by
\begin{eqnarray}
  \ln Z^{\mathrm{ws}} &=&
  \alpha
  \left[ -2p(x_m) \left(\ln \frac{\sqrt{2\pi \alpha}}{\lambda^2 n_b}+b_0\right)
    +\int_{1/x_m}^{x_m} \frac{(1+x)^4}{x^3}\,\ln\frac{(x+1)^4}{x^2}\,dx
    \right]
  \nonumber\\
  &&
  -2\sqrt{\frac{4\pi\alpha}{n_b}}\, x_m \left(1+\frac{1}{x_m}\right)^2
  \beta\gamma_{\text{hard}}
  +O(1)
  \label{eq:free-energy-fixed-shape-full-surface}
\end{eqnarray}
We notice that the free energy has again a nonextensive term
proportional to $\alpha\ln {\alpha}$, but, once again, it can be
  cured by choosing the constant $b_0$ as $b_0=-\ln( M
  x_m)+\text{constant}$. The perimeter correction,
  $2\mathcal{C}_R\beta\gamma_{\text{hard}}$,  proportional to
  $\sqrt{\alpha}$, has the same form as for the half surface case, with
  equal contributions from each boundary at $x=1/x_m$ and
  $x=x_m$. Once again, there is no $\ln \alpha$ correction in agreement
  with the general theory of Ref.~\cite{Jancovici94,Jancovici96} and
  the fact that the Euler characteristic of this manifold is $\chi=0$.

\subsubsection{Density}

The density is now given by
\begin{equation}
  n^{\mathrm{ws}}(x)=\sum_{k=0}^{N-1}
  \frac{x^{2k-N+1}\,e^{-\alpha h(x)}}{\tilde{\mathcal{B}}_N(k)}
\end{equation}
Due to the fact that the asymptotic behavior of
$\tilde{\mathcal{B}}_N(k)$ is almost the same as the one of
$\mathcal{B}_N(k')$ with $k'=|k-\frac{N}{2}|$, the behavior of the
density turn out to be the same as for the half surface case, in the
thermodynamic limit $\alpha\to\infty$, $x_m$ fixed,
\begin{equation}
  n(x)=n_b\,,
  \qquad\text{in the bulk, ie., when $x-x_m$ and $x-\frac{1}{x_m}$ are
    of order 1.}
\end{equation}
And, close to the boundaries, $x\to x_b$ with $x_b=x_m$ or $x_b=1/x_m$,
\begin{equation}
  \label{eq:density-border-xb}
  n(x)=\frac{2 n_b}{\sqrt{\pi}}
  \int_0^{+\infty}
  \frac{\exp\left[-\left(t-\sqrt{\frac{\alpha p'(x_b)}{x_b}}|x-x_b|)
        \right)^2\right]}{1+\erf(t)}
    \,dt\,,
    \quad
    \text{for $x_b-x$ of order }\frac{1}{\sqrt{\alpha}}
    \,.
\end{equation}
If the result is expressed in terms of the geodesic distance $s$ to the
border, we recover, once again, the result of the OCP in a flat
space near a hard wall~(\ref{eq:density-border-flat}).


\subsection{The 2dOCP on the half surface with potential 
$-\ln(|z-z'|/\sqrt{|zz'|})$}

\subsubsection{Partition function}

In this case, we have $N=\alpha p(x_m)$. Following similar calculations
to the ones of the previous cases, we find that the partition function, at
$\beta q^2=2$, is
\begin{equation}
  Z^{\overline{\mathrm{hs}}}=Z_0^{\overline{\mathrm{hs}}}
  e^{-\beta F_0^{\overline{\mathrm{hs}}}}
\end{equation}
with 
\begin{equation}
  -\beta F_0^{\overline{\mathrm{hs}}}
  = \alpha^2 p(x_m) h(x_m) - p(x_m)^2\ln x_m
  +\int_1^{x_m} \frac{\left[p(x)\right]^2}{x}\,dx
  -Nb_0
\end{equation}
and 
\begin{equation}
  Z_0^{\overline{\mathrm{hs}}}=\prod_{k=0}^{N-1} \hat{\mathcal{B}}_N(k)
\end{equation}
with
\begin{equation}
  \hat{\mathcal{B}}_N(k)=\frac{\alpha}{n_b}\int_1^{x_m}
  x^{2k+1}e^{-\alpha h(x)}\,dx
\end{equation}

\subsubsection{Thermodynamic limit $R\to\infty$, $x_m\to\infty$, and fixed $M$}

The asymptotic expansion of $\hat{\mathcal{B}}_{N}(k)$ is obtained from
equation~(\ref{eq:asympt-tildeBN}) replacing $k'$ by $k$ and
considering only the case $k>0$. As explained in
appendix~\ref{app:gamma}, the main difference with the other half
surface case (section \ref{sec:half-surface-1}), is an additional term
$\hx_k$ in each factor of the partition function and the additional term
$(3/(8k))$ in the expansion~(\ref{eq:asympt-tildeBN}). Therefore, the
partition function can be obtained from the one of the half surface
with potential $-\ln|z-z'|$ by adding the terms
\begin{eqnarray}
  S_4^{\overline{\mathrm{hs}}}&=&\sum_{k=0}^{N-1} \ln \hx_k\,,\\
  S_5^{\overline{\mathrm{hs}}}
  &=&\sum_{k=1}^{N-1} \frac{3}{8k}=\frac{3}{8}\ln
  N+O(1)=\frac{3}{4}\ln x_m +O(1)\,.
\end{eqnarray}
Using Euler-McLaurin expansion, we have
\begin{eqnarray}
  S_4^{\overline{\mathrm{hs}}}&=&\sum_{k=0}^{N} \ln \hx_k -\ln x_m
  \nonumber\\
  &=&\int_1^{x_m} \alpha p'(x)\ln x\,dx +\frac{1}{2}\ln x_m - \ln x_m+ O(1)
  \nonumber\\
  &=& 
  \alpha p(x_m)\ln x_m -\alpha \int_1^{x_m} \frac{p(x)}{x}\,dx 
  -\frac{1}{2} \ln x_m+O(1)
  \nonumber\\
  &=&
  \alpha p(x_m)\ln x_m - \frac{1}{2}\alpha h(x_m) -\frac{1}{2}\ln x_m+ O(1)
\,,
\end{eqnarray}
where we used the property~(\ref{eq:p-h-deriv}).
Finally, 
\begin{eqnarray}
  \ln Z^{\overline{\mathrm{hs}}}&=&\alpha x_m^2 \ln x_m+ N \left(b_0+
  \ln\frac{\sqrt{2\pi\alpha}}{\lambda^2 n_b}\right)
  -\frac{\alpha}{2} x_m^2 
  +4\alpha x_m (2 \ln x_m -1)
  \nonumber\\
  &&
  - \mathcal{C}_R\,\beta \gamma_{\text{hard}}
  +6 \alpha (\ln x_m)^2+14 \alpha \ln x_m +\frac{1}{12}\ln x_m
  +O(1)\,.
\end{eqnarray}
The result is one-half of the one for the full surface, $\ln
Z^{\mathrm{ws}}$, as it might be expected.

\subsubsection{Thermodynamic limit at fixed shape: $\alpha\to\infty$ and
  $x_m$ fixed}

For this case, the asymptotics of $\hat{\mathcal{B}}_N$ are very
similar to those of $\mathcal{B}_N$ from
equation~(\ref{eq:BN-fixed-shape})
\begin{equation}
  \hat{\mathcal{B}}_N(k)\sim \hx_k \mathcal{B}_N(k)
  \,.
\end{equation}
Therefore, the only difference from the calculations of the half surface case
with potential $-\ln|z-z'|+\text{constant}$, and this case, is the sum
\begin{equation}
  S_4^{\overline{\mathrm{hs}},\mathrm{fixed}}=\sum_{k=0}^{N-1} \ln \hx_k\,.
\end{equation}
We have
\begin{eqnarray}
  S_4^{\overline{\mathrm{hs}},\mathrm{fixed}}
  &=&\int_1^{x_m} \alpha p'(x)\ln x\,dx + O(1)
  \nonumber\\
  &=&
  \alpha p(x_m)\ln x_m - \frac{1}{2}\alpha h(x_m)+ O(1)
  \,.
\end{eqnarray}
Here, the term $k=N$ and the remainder of the Euler-McLaurin expansion
give corrections of order $O(\alpha^0)=O(1)$, as opposed to the
previous section where they gave contributions of order $O(\ln x_m)$.

Finally, we find
\begin{eqnarray}
  \ln Z^{\overline{\mathrm{hs}}} &=& \alpha \left[ p(x_m)\left(
    \frac{1}{2}\ln\frac{\sqrt{2\alpha \pi}}{n_b}+b_0 \right) +
    \int_{1}^{\infty}
    \frac{(1+x)^4}{x^3}\ln\frac{(1+x)^4}{x^2}\,dx\right]
  \nonumber\\ && -\sqrt{\frac{4\pi\alpha}{n_b}}
  \left[x_m\left(1+\frac{1}{x_m}\right)+4\right]\beta\gamma_{\text{hard}}
  +O(1)\,.
\end{eqnarray}
The bulk free energy, proportional to $\alpha$, plus the nonextensive
term proportional $\alpha\ln\alpha$, are one-half the ones from
equation~(\ref{eq:free-energy-fixed-shape-full-surface}) for the full
surface case, as expected. The perimeter contribution, proportional to
$\sqrt{\alpha}$ is again the same as in all the previous cases of
thermodynamic limit at fixed shape, i.e.~a contribution $\beta
\gamma_{\text{hard}} \mathcal{C}_b$ for each boundary at $x_b=x_m$ and
at $x_b=1$ with perimeter $\mathcal{C}_b$~(\ref{eq:perimeter}). Once
again, there is no $\ln \alpha$ correction in agreement with the fact
that the Euler characteristic of this manifold is $\chi=0$.

\subsection{The grounded horizon case}

\subsubsection{Grand canonical partition function}

In order to find the partition function for the system in the half
space, with a metallic grounded boundary at $x=1$, when the charges
interacting through the pair potential of Eq.~(\ref{ghgreen}) it is
convenient to work in the grand canonical ensemble instead, and use
the techniques developed in Refs.~\cite{Forrester85,Jancovici96}.  We
consider a system with a fixed background density $\rho_b$. The
fugacity $\tilde{\zeta}=e^{\beta \mu}/\lambda^2$, where $\mu$ is the
chemical potential, controls the average number of particles $\langle
N\rangle$, and in general the system is nonneutral $\langle N\rangle
\neq N_b$, where $N_b=\alpha p(x_m)$. The excess charge is expected to
be found near the boundaries at $x=1$ and $x=x_m$, while in the bulk
the system is expected to be locally neutral. In order to avoid the
collapse of a particle into the metallic boundary, due to its
attraction to the image charges, we confine the particles to be in a
``disk'' domain $\tilde{\Omega}_R$, where $x\in[1+w,x_m]$. We
introduced a small gap $w$ between the metallic boundary and the
domain containing the particles, the geodesic width of this gap is
$W=\sqrt{\alpha p'(1)/(2\pi n_b)}\,w$. On the other hand, for
simplicity, we consider that the fixed background extends up to the
metallic boundary.

In the potential energy of the system~(\ref{eq:hamiltonian-gen}) we
should add the self energy of each particle, that is due to the fact
that each particle polarizes the metallic boundary, creating an
induced surface charge density. This self energy is $\frac{q^2}{2}\ln
[|x^2-1|M/2L]$, where the constant $\ln (M/2L)$ has been added to
recover, in the limit $M\to0$, the self energy of a charged particle
near a plane grounded wall in flat space.

The grand partition function, when $\beta
q^2=2$, is
\begin{equation}
\Xi=e^{-\beta F_0^{\text{gh}}} \left[1+\sum_{N=1}^{\infty}
  \frac{\zeta^N}{N!}  \int \prod_{i=1}^N dS_i
  \prod_{i<j}\left|\frac{z_i-z_j}{1-z_i \bar{z}_j}\right|^2
  \prod_{i=1}^{N} \left| |z_i|^2-1\right|^{-1} 
  \prod_{i=1}^{N} e^{-\alpha [ h(x_i)-2N_b \ln x_i ]}
  \right]
\end{equation}
where for $N=1$ the product $\prod_{i<j}$ must be replaced by 1. The
domain of integration for each particle is $\tilde{\Omega}_R$. We
have defined a rescaled fugacity $\zeta=2L\tilde{\zeta }/M$ and
\begin{equation}
  -\beta F_0^{\text{gh}}=\alpha N_b h(x_m) - N_b^2\ln x_m 
  -\alpha^2 \int_{1}^{x_m} \frac{[p(x)]^2}{x}\,dx
\end{equation}
which is very similar to $F_0^{\text{hs}}$, except that here
$N_b=\alpha p(x_m)$ is not equal to $N$ the number of particles.

Let us define a set of reduced complex coordinates $u_i=z_i$
and its corresponding images $u_i^*=1/\bar{z}_i$. By using Cauchy
identity 
\begin{eqnarray} \label{Cauchy}
\det
\left(
\frac{1}{u_i-u_j^*}
\right)_{(i,j)\in\{1,\cdots,N\}^2}
=
(-1)^{N(N-1)/2}\:
\frac{\prod_{i<j} (u_i-u_j)(u^*_i-u^*_j)}{\prod_{i,j} (u_i-u_j^*)}
\end{eqnarray}
the particle-particle interaction and self energy terms can be
cast into the form
\begin{eqnarray}
\prod_{i<j}\left|
\frac{z_i-z_j}{1-z_i \bar{z}_j}
\right|^2
\prod_{i=1}^N \left(|z_i|^2-1\right)^{-1}
=(-1)^{N}
\det
\left(
\frac{1}{1-z_i \bar{z}_j}
\right)_{(i,j)\in\{1,\cdots,N\}^2}
\,.
\end{eqnarray}
The grand canonical partition function is then
\begin{eqnarray}
\label{eq:GrandPart-prelim}
\Xi= e^{-\beta F_0^{\text{gh}}}\left[1+\sum_{N=1}^{\infty} \frac{1}{N!}
\int \prod_{i=1}^N dS_i 
\prod_{i=1}^{N} \left[-\zeta(x_i)\right]\,\det
\left(
\frac{1}{1-z_i \bar{z}_j}
\right)
\right]
\,,
\end{eqnarray}
with $\zeta(x)=\zeta  e^{-\alpha[h(x)-2N_b \ln x]}$.
We shall now recall how this expression can be reduced to a Fredholm
determinant~\cite{Forrester85}. Let us consider the Gaussian partition
function
\begin{equation}
\label{eq:part-fun-Grass-libre}
Z_0=\int {\cal D}\psi {\cal D}\bar{\psi} \,\exp\left[\int \bar{\psi}(\qq)
A^{-1}(z,\bar{z}') \psi(\qq')\, dS\, dS' \right]
\,.
\end{equation}
The fields $\psi$ and $\bar{\psi}$ are anticommuting Grassmann
variables.  The Gaussian measure in~(\ref{eq:part-fun-Grass-libre}) is
chosen such that its covariance is equal to
\begin{equation}
\left<\bar{\psi}(\qq_i)\psi(\qq_j)\right>
=
A(z_i,\bar{z}_j)=\frac{1}{1-z_i \bar{z}_j}
\end{equation}
where $\langle\ldots\rangle$ denotes an average taken with the Gaussian
weight of (\ref{eq:part-fun-Grass-libre}). By construction we have
\begin{equation} \label{Z_0}
Z_0=\det(A^{-1})
\end{equation}
Let us now consider the following partition function
\begin{equation}
Z=\int {\cal D}\psi {\cal D}\bar{\psi} \exp\left[\int \bar{\psi}(\qq)
A^{-1}(z,\bar{z}') \psi(\qq') dS dS' -\int \zeta(x) 
\bar{\psi}(\qq)\psi(\qq) \,dS \right]
\end{equation}
which is equal to
\begin{equation}
Z=\det(A^{-1}-\zeta)
\end{equation}
and then
\begin{equation} \label{Z/Z_0}
\frac{Z}{Z_0}=\det[A(A^{-1}-\zeta)]=\det(1+K)
\end{equation}
where $K$ is an integral operator (with integration measure $dS$)
with kernel
\begin{equation} \label{K}
K(\qq,\qq')=-\zeta(x')\, A(z,\bar{z}')=
-\frac{\zeta(x')}{1-z\bar{z}'}
\,.
\end{equation}

Expanding the ratio $Z/Z_0$ in 
powers of $\zeta$ we have
\begin{equation}
\label{eq:expans-ZZ0}
\frac{Z}{Z_0}=
1+
\sum_{N=1}^{\infty}
\frac{1}{N!}
\int \prod_{i=1}^N dS_i
(-1)^{N}\prod_{i=1}^N
\zeta(x_i)
\left<\bar{\psi}(\qq_1)\psi(\qq_1)\cdots
\bar{\psi}(\qq_N)\psi(\qq_N)\right>
\end{equation}
Now, using Wick theorem for anticommuting variables~\cite{ZinnJustin},
we find that
\begin{equation}
\label{eq:WickFerms}
\left<\bar{\psi}(\qq_1)\psi(\qq_1)\cdots
\bar{\psi}(\qq_N)\psi(\qq_N)\right>
=\det A(z_i,\bar{z}_j)=\det\left(\frac{1}{1-z_i \bar{z}_j}\right)
\end{equation}
Comparing equations~(\ref{eq:expans-ZZ0})
and~(\ref{eq:GrandPart-prelim}) with the help of
equation~(\ref{eq:WickFerms}) we conclude that
\begin{equation}
\label{eq:Xi-det}
\Xi=e^{-\beta F_0^{\text{gh}}}\,\frac{Z}{Z_0}=e^{-\beta F_0^{\text{gh}}}\det(1+K)
\end{equation}

The problem of computing the grand canonical partition function has
been reduced to finding the eigenvalues $\lambda$ of the operator
$K$. The eigenvalue problem for $K$ reads
\begin{equation}
\label{eq:vpK}
-\int_{\tilde{\Omega}_R} 
\frac{\zeta(x')}
{ 1-z\bar{z}'}\,
\Phi(x',\varphi') dS'
=
\lambda \Phi(x,\varphi)
\end{equation}
For $\lambda\neq 0$ we notice from equation~(\ref{eq:vpK}) that
$\Phi(x,\varphi)=\Phi(z)$ is an analytical function of
$z=xe^{i\varphi}$ in the region $|z|>1$. Because of the circular
symmetry, it is natural to try $\Phi(z)=\Phi_{\ell}(z)=z^{-\ell}$ with
$\ell\ge 1$ a positive integer. Expanding
\begin{equation}
\frac{1}{1-z\bar{z}'}=
-\sum_{n=1}^{\infty}\left(z\bar{z}'\right)^{-n}
\end{equation}
and replacing $\Phi_{\ell}(z)=z^{-\ell}$ in equation~(\ref{eq:vpK}), we
show that $\Phi_{\ell}$ is indeed an eigenfunction of $K$ with
eigenvalue
\begin{equation}
\label{eq:lambda-vp-de-K}
\lambda_{\ell}=
\zeta \mathcal{B}_{N_b}^{\text{gh}}(N_b-\ell)
\end{equation}
where 
\begin{equation}
  \label{eq:BNgh}
  \mathcal{B}_{N_b}^{\text{gh}}(k)=\frac{\alpha}{n_b}\int_{1+w}^{x_m}
  x^{2k} e^{- \alpha h(x)}\,p'(x)\,dx
\end{equation}
which is very similar to $\mathcal{B}_N$ defined in Eq.~(\ref{gamma}),
except for the small gap $w$ in the lower limit of integration. So, we
arrive to the result for the grand potential
\begin{equation}
\label{eq:grand-potential-somme}
\beta\Omega = -\ln\Xi 
=
\beta F_0^{\text{gh}}
-
\sum_{\ell=1}^{\infty}
\ln\left[
1+\zeta {\cal B}_{N_b}^{\text{gh}}(N_b-\ell)
\right]\,.
\end{equation}

\subsubsection{Thermodynamic limit at fixed shape: $\alpha\to\infty$ and $x_m$
  fixed}

Let us define $k=N_b-\ell$ for $\ell\in\mathbb{N}^{*}$, thus $k$ is positive,
then negative when $\ell$ increases. Therefore, it is convenient to
split the sum~(\ref{eq:grand-potential-somme}) in $\ln\Xi$ into two
parts
\begin{eqnarray}
  \label{eq:S6gh}
  S_{6}^{\text{gh,fixed}}
  &=& \sum_{k=-\infty}^{-1}
  \ln[1+\zeta \mathcal{B}_{N_b}^{\text{gh}}(k)]
  \\
  \label{eq:S7gh}
  S_{7}^{\text{gh,fixed}}&=&\sum_{k=0}^{N_b-1}
  \ln[1+\zeta \mathcal{B}_{N_b}^{\text{gh}}(k)]\,.
\end{eqnarray}

The asymptotic behavior of $\mathcal{B}_{N_b}^{\text{gh}}(k)$ when
$\alpha\to\infty$ can be directly deduced from the one of
$\mathcal{B}_N$ found in appendix~\ref{app:gamma},
Eq.~(\ref{eq:BN-fixed-shape}), taking into account the small gap $w$
near the boundary at $x=1+w$. When $k<0$, we have $\hx_k<1$, then we
notice that $\epsilon_{k,1}$ defined in~(\ref{eq:epsilon_k1}) is
negative, and that the relevant contributions to the sum
$S_6^{\text{gh,fixed}}$ are obtained when $k$ is close to 0, more precisely
$k$ of order $O(\sqrt{N_b})$. So, we expand $\hx_k$ around $\hx_k=1$ up to
order $(\hx_k-1)^2$ in the exponential term
$e^{-\alpha[h(\hx_k)-2p(\hx_k)\ln \hx_k]}$ from
Eq.~(\ref{eq:BN-fixed-shape}). Then, we have, for $k<0$ of order
$O(\sqrt{N_b})$,
\begin{equation}
  \label{eq:BNgh-asympt}
  \mathcal{B}_{N_b}^{\text{gh}}(k)=
  \frac{\sqrt{\alpha \pi p'(1)}}{2 n_b} 
  e^{\alpha p'(1)\,(1-\hx_k)^2}
  \erfc[\sqrt{\alpha p'(1)}\,(1+w-\hx_k)]
\end{equation}
where $\erfc(u)=1-\erf(u)$ is the complementary error function. Then,
up to corrections of order $O(1)$, the sum $S_6^{\text{gh,fixed}}$ can be
transformed into an integral over the variable $t=\sqrt{\alpha
  p'(1)}\,(1-\hx_k)$, to find
\begin{equation}
  S_6^{\text{gh,fixed}}=\sqrt{\alpha p'(1)}\int_{0}^{\infty}
  \ln\left[
    1+\frac{\zeta \sqrt{\alpha\pi p'(1)}}{2n_b}
    e^{t^2}\,\erfc\left(t+\sqrt{2\pi n_b} W\right)
    \right]\,dt+O(1)
  \,.
\end{equation}
Let $\mathcal{C}_1=\sqrt{2\pi\alpha p'(1)/n_b}$, be total
length of the boundary at $x=1$. We notice that 
\begin{equation}
  \zeta\frac{\sqrt{\alpha \pi p'(1)}}{2n_b}
  =\frac{\zeta \mathcal{C}_{1}}{\sqrt{2n_b}}
  =\frac{2\tilde{\zeta}L}{\sqrt{2n_b}}
  \frac{\mathcal{C}_{1}}{M}
\end{equation}
is fixed and of order $O(1)$ in the limit $M\to\infty$, since in the
fixed shape limit $\mathcal{C}_{1}/M$ is fixed. Therefore
$S_6^{\text{gh,fixed}}$ gives a contribution proportional to the
perimeter $\mathcal{C}_1$.

For $S_7^{\text{gh,fixed}}$, we define
\begin{equation}
\tilde{\epsilon}_{k,1}=\sqrt{\alpha p'(1)}\,(1+w-\hx_k)  
\,,
\end{equation}
and we write
\begin{eqnarray}
  S_7^{\text{gh,fixed}}&=&\sum_{k=0}^{N_b-1} \ln \left[
    1+\frac{\zeta\sqrt{\alpha \pi \hx_k p'(\hx_k)}}{2n_b}
    e^{-\alpha[h(\hx_k)-2p(\hx_k)\ln \hx_k]}
    \left[\erf(\tilde{\epsilon}_{k,1})+\erf({\epsilon}_{k,m})
      \right]
    \right]
  \nonumber\\
  &=& S_8^{\text{gh,fixed}} + S_{1}^{\text{hs,fixed}}
    +S_{2}^{\text{hs,fixed}}+N_b\ln \frac{\zeta \sqrt{\alpha\pi}}{n_b}
  \end{eqnarray}
where
\begin{equation}
  S_8^{\text{gh,fixed}}=\sum_{k=0}^{N_b-1}
  \ln\left[
    \frac{n_b e^{\alpha[h(\hx_k)-2p(\hx_k)\ln \hx_k]}}{\zeta\sqrt{\alpha\pi
          \hx_k p'(\hx_k)}}
      +\frac{1}{2}\left[
        \erf(\tilde{\epsilon}_{k,1})+\erf(\epsilon_{k,m})
        \right]
    \right]
\end{equation}
and we see that the sums $S_{1}^{\text{hs,fixed}}$ and
$S_{2}^{\text{hs,fixed}}$ reappear. These are defined in
equations~(\ref{eq:S1-hs-fixed}) and~(\ref{eq:S2-hs-fixed}) and
computed in~(\ref{eq:S1-hs-fixed-asympt})
and~(\ref{eq:S2-hs-fixed-asympt}). In a similar way to
$S_6^{\text{gh,fixed}}$, $S_8^{\text{gh,fixed}}$ gives only boundary
contributions when $k$ is close to 0, of order $\sqrt{N_b}$ (grounded
boundary at $x=1$) and when $k$ is close to $N_b$ with $N_b-k$ of
order $\sqrt{N_b}$ (boundary at $x=x_m$). We have,
\begin{eqnarray}
  S_8^{\text{gh,fixed}}&=&
  \sqrt{\alpha p'(1)}
  \int_{0}^{\infty}
  \ln\left[
    \frac{n_b e^{-t^2}}{\zeta \sqrt{\alpha \pi p'(1)}}
      +\frac{1}{2}\left[\erf(t-\sqrt{2\pi n_b} W)+1\right]
    \right]\,dt
    \nonumber\\
    &&
    +
  \sqrt{\alpha x_m p'(x_m)}
  \int_{0}^{\infty}
  \ln\left[
    \frac{\erf(t)+1}{2}
    \right]\,dt
\end{eqnarray}
Let us introduce again the perimeter of the outer boundary at $x=x_m$,
$\mathcal{C}_{R}=\sqrt{2\pi\alpha x_m p'(x_m)/n_b}$. Putting together
all terms, we finally have
\begin{eqnarray}
  \ln \Xi &=&
  -N_b \beta\omega_{B}+\frac{\alpha}{2}\left[h(x_m)-2p(x_m)\ln
    x_m\right]
  +\alpha \int_{1}^{x_m} \frac{(1+x)^4}{x^3}\ln\frac{(1+x)^4}{x^2}\,dx
  \nonumber\\
  &&
  -\mathcal{C}_1 \beta\gamma_{\text{metal}}
  -\mathcal{C}_{R} \beta\gamma_{\text{hard}}
  +O(1)
\end{eqnarray}
where
\begin{equation}
  \beta\omega_{B} = -\ln \frac{2\pi\tilde{\zeta}L}{\sqrt{2n_b}}
\end{equation}
is the bulk grand potential per particle of the OCP near a plane
metallic wall in the flat space. The surface (perimeter) tensions
$\gamma_{\text{metal}}$ and $\gamma_{\text{hard}}$ associated to each
boundary (metallic at $x_b=1$, and hard wall at $x_b=x_m$) are given
by
\begin{eqnarray}
  \beta  \gamma_{\text{metal}}&=&-  \sqrt{\frac{n_b}{2\pi}}
  \int_{0}^{\infty}
  \ln\left[
    1+\frac{\zeta \sqrt{\alpha\pi x_b p'(x_b)}}{2n_b}
    e^{t^2}\,\erfc\left(t+\sqrt{2\pi n_b} W\right)
    \right]\,dt
  \nonumber\\
  &&
-  \sqrt{\frac{n_b}{2\pi}}
  \int_{0}^{\infty}
  \ln\left[
    \frac{n_b e^{-t^2}}{\zeta \sqrt{\alpha \pi p'(x_b) x_b}}
      +\frac{1}{2}\left[\erf(t-\sqrt{2\pi n_b} W)+1\right]
    \right]
  \,dt
\end{eqnarray}
with $x_b=1$, and (\ref{app:gam}) for $\beta \gamma_{\text{hard}}$.

Notice, once again, that the combination
\begin{equation}
  \frac{\zeta \sqrt{\alpha \pi x_b p'(x_b)}}{2n_b}
  =\frac{2\tilde{\zeta}L}{\sqrt{2 n_b}} \frac{\mathcal{C}_b}{M}
\end{equation}
is finite in this fixed shape limit, since the perimeter
$\mathcal{C}_{b}$ of the boundary at $x_b$ scales as $M$. Up to a
rescaling of the fugacity $\tilde{\zeta}$ to absorb the factor
$\mathcal{C}_b/M$, the surface tension near the metallic boundary
$\gamma_{\text{metal}}$ is the same as the one found in
Ref.~\cite{Jancovici96} in flat space. It is also similar to the
one found in Ref.~\cite{Forrester85} with a small difference due to
the fact that in that reference the background does not extend up to
the metallic boundary, but has also a small gap near the boundary. 

There is no $\ln \alpha$ correction in the grand potential in
agreement with the fact that the Euler characteristic of the manifold
is $\chi=0$.

Let us decompose $\ln \Xi$ into its bulk and perimeter parts,
\begin{equation}
  \label{eq:lnXi-bulk-surface}
  \ln \Xi = -\beta \Omega_b^{\text{gh}}   
  -\mathcal{C}_1 \beta\gamma_{\text{metal}}
  -\mathcal{C}_{R} \beta\gamma_{\text{hard}}
  +O(1)
\end{equation}
with the bulk grand potential $\Omega_{b}^{\text{gh}}$ given by
\begin{equation}
  -\beta \Omega_b^{\text{gh}}   =
  -N_b \beta\omega_{B}+\frac{\alpha}{2}\left[h(x_m)-2p(x_m)\ln
    x_m\right]
  +\alpha \int_{1}^{x_m} \frac{(1+x)^4}{x^3}\ln\frac{(1+x)^4}{x^2}\,dx
  \,.
\end{equation}
The average number of particles is given by the usual thermodynamic
relation $\langle N \rangle = \zeta \partial(\ln\Xi)/\partial
\zeta$. Following~(\ref{eq:lnXi-bulk-surface}), it can be decomposed
into bulk and perimeter contributions,
\begin{equation}
  \langle N \rangle = N_b - \mathcal{C}_1\zeta \frac{\partial \beta
    \gamma_{\text{metal}}}{\partial \zeta}
  \,.
\end{equation}
The boundary at $x=x_m$ does not contribute because
$\gamma_{\text{hard}}$ does not depend on the fugacity. From this
equation, we can deduce the perimeter linear charge density $\sigma$
which accumulates near the metallic boundary
\begin{equation}
  \sigma = - \zeta \frac{\partial \beta
    \gamma_{\text{metal}}}{\partial \zeta}
  \,.
\end{equation}
We can also notice that the bulk Helmoltz free energy
$F_{b}^{\text{gh}} = \Omega_{b}^{\text{gh}} + \mu N_b$ is the same as
for the half surface, with Coulomb potential $G^{\text{hs}}$, given in
(\ref{eq:free-energy-fixed-shape}).

\subsubsection{Thermodynamic limit $R\to\infty$, $x_m\to \infty$, and
  fixed $M$}

This limit is of restricted interest, since the metallic boundary
perimeter remains of order $O(1)$, we expect to find the same
thermodynamic quantities as in the half surface case with hard wall
``horizon'' boundary up to order $O(\ln x_m)$. This is indeed the
case: let us split $\ln \Xi$ into two sums $S_6^{\text{gh}}$ and
$S_7^{\text{gh}}$ as in~(\ref{eq:S6gh}) and~(\ref{eq:S7gh}). For
$k<0$, the asymptotic expansion of $\mathcal{B}_{N_b}(k)$ derived in
appendix~\ref{app:gamma} should be revised, because the absolute
maximum of the integrand is obtained for values of the variable of
integration outside the domain of integration. Within the domain of
integration the maximum value of the integrand in~(\ref{eq:BNgh}) is
obtained when $x=1+w$. Expanding the integrand around that value, we
obtain to first order, for large $|k|$,
\begin{equation}
  \mathcal{B}_{N_b}^{\text{gh}}(k)\sim\frac{\alpha p'(1+w)}{2n_b
    |k|}e^{-2w |k|}
  \,.
\end{equation}
Then
\begin{eqnarray}
  S_6^{\text{gh}}&=&\sum_{k=-\infty}^{0} \ln\left[1+\zeta
  \mathcal{B}_{N_b}^{\text{gh}}(k)\right]
  \nonumber\\
  &=& \int_{0}^{\infty} dk \ln \left[1+\zeta
    \frac{\alpha p'(1+w)}{2n_b
    |k|}e^{-2w |k|}\right] + O(1)
  \nonumber\\
  &=& O(1)\,,
\end{eqnarray}
does not contribute to the result at orders greater than $O(1)$. For
the other sum, we have
\begin{eqnarray}
  S_7^{\text{gh}}&=&\sum_{k=0}^{N_b} \ln\left[\zeta
    \mathcal{B}_{N_b}^{\text{gh}}(k) \right]
  +\sum_{k=0}^{N_b} \ln \left[ 1 +
    \frac{1}{\zeta \mathcal{B}_{N_b}^{\text{gh}}(k)}
    \right]
  \nonumber\\
  &=&
  \sum_{k=0}^{N_b} \ln\left[\zeta
    \mathcal{B}_{N_b}^{\text{gh}}(k) \right]
  +O(1)\,.
\end{eqnarray}
The second sum is indeed $O(1)$, because $1/[\zeta
  \mathcal{B}_{N_b}^{\text{gh}}(k)]$ has a fast exponential decay for
large $k$, therefore the sum can be converted into an finite [order
$O(1)$] integral over the variable $k$. 

Now, since the asymptotic behavior of
$\mathcal{B}_{N_b}^{\text{gh}}(k)$, for $k>0$ and large, is
essentially the same as the one for $\mathcal{B}_{N_b}(k)$, we
immediately find, up to $O(1)$ corrections,
\begin{equation}
  \ln \Xi = \beta \mu N_b + \ln Z^{\text{hs}}
  +O(1)
\end{equation}
where $\ln Z^{\text{hs}}$ is minus the free energy in the half surface case with
hard wall boundary, given by~(\ref{eq:Fhs-xm-infinity}).

\subsubsection{The one-body density}

As usual one can compute the density by doing a functional derivative
of the grand potential with respect to a position-dependent fugacity
$\zeta(\qq)$
\begin{equation}
\label{eq:n-funct-deriv}
n^{\text{gh}}(\qq)=
\zeta(\qq)\frac{\delta\ln\Xi}{\delta \zeta(\qq)}
\,.
\end{equation}
For the present case of a curved space, we shall understand the
functional derivative with the rule $\frac{\delta \zeta(\qq')}{\delta
  \zeta(\qq)}=\delta(\qq,\qq')$ where
$\delta(\qq,\qq')=\delta(x-x')\delta(\varphi-\varphi')/\sqrt{g}$ is the
Dirac distribution on the curved surface.

Using a Dirac-like notation, one can formally write
\begin{equation}
\ln\Xi=\mbox{Tr} \ln(1+K)-\beta F_0^{\text{gh}}=
\int \left<\qq\left|
\ln(1-\zeta(\qq)A)\right|\qq\right>
\,dS
-\beta F_0^{\text{gh}}
\end{equation}
Then, doing the functional derivative~(\ref{eq:n-funct-deriv}), one
obtains
\begin{equation}
n^{\text{gh}}(\qq)=
\zeta\left<\qq\left| (1+K)^{-1}(-A) \right|\qq\right> =
\zeta G(\qq,\qq)
\end{equation}
where we have defined $G(\qq,\qq')$ by
$G=(1+K)^{-1}(-A)$. More explicitly, $G$ is the solution
of $(1+K)G=-A$, that is
\begin{equation}
\label{eq:eq-Green-function}
G(\qq,\qq') - \int_{\tilde{\Omega}_R}
\zeta(x'')\frac{G(\qq'',\qq')}{1-z\bar{z}''}
\, dS'' = -\frac{1}{1-z\bar{z}'}
\,.
\end{equation}
From this integral equation, one can see that ${G}(\qq,\qq')$ is
an analytical function of $z$ in the region $|z|>1$. Then, we look for
a solution in the form of a Laurent series
\begin{equation}
{G}(\qq,\qq')=\sum_{\ell=1}^{\infty} a_{\ell}(\rr') z^{-\ell}
\end{equation}
into equation~(\ref{eq:eq-Green-function}) yields
\begin{equation}
\label{eq:solution-G}
{G}(\qq,\qq')=
\sum_{\ell=1}^{\infty}
\frac{\left(z\bar{z}'\right)^{-\ell}}{1+\lambda_\ell}
\,.
\end{equation}
Recalling that $\lambda_{\ell}=\zeta\mathcal{B}_{N}^{\text{gh}}(N_b-\ell)$,
the density is given by
\begin{equation}
\label{eq:densite-somme}
n^{\text{gh}}(x)=
\zeta
\sum_{k=-\infty}^{N_b-1}
\frac{x^{2k}e^{-\alpha h(x)}}{1+\zeta \mathcal{B}_{N}^{\text{gh}}(k)}
\end{equation}

\subsubsection{Density in the thermodynamic limit at fixed shape
  $\alpha\to\infty$ and $x_m$ fixed.}

Using the asymptotic behavior~(\ref{eq:BN-fixed-shape}) of
$\mathcal{B}_N^{\text{gh}}$, we have
\begin{equation}
  n^{\text{gh}}(x)=\zeta
  \sum_{k=-\infty}^{N_b}
  \frac{\exp\left(
      -\alpha[h(x)-2p(\hx_k)\ln x - h(\hx_k) + 2 p(\hx_k)\ln \hx_k]
      \right)}{
    e^{\alpha[h(\hx_k)-2p(\hx_k)\ln \hx_k]}
    +\frac{\zeta \sqrt{\alpha \pi \hx_k p'(\hx_k)}}{2n_b}
    \left[ \erf(\tilde{\epsilon}_{k,1})+\erf(\epsilon_{k,m})
      \right]}
\,.
\end{equation}
Once again, this sum can be evaluated using Laplace method. The
exponential in the numerator presents a peaked maximum for $k$ such
that $\hx_k=x$. Expanding the argument of the exponential around its
maximum, we have
\begin{equation}
  n^{\text{gh}}(x)=\zeta
  \sum_{k=-\infty}^{N_b}
  \frac{e^{-\alpha p'(x) (x-\hx_k)^2/x}}{
    e^{\alpha[h(\hx_k)-2p(\hx_k)\ln \hx_k]}
    +\frac{\zeta \sqrt{\alpha \pi \hx_k p'(\hx_k)}}{2n_b}
    \left[ \erf(\tilde{\epsilon}_{k,1})+\erf(\epsilon_{k,m})
      \right]}
\,.
\end{equation}
Now, three cases has to be considered, depending on the value of $x$.

If $x$ is in the bulk, {\sl i.e.\/}~$x-1$ and $x_m-x$ of order 1, the
exponential term in denominator vanishes in the limit $\alpha\to \infty$,
and we end up with an expression which is essentially the same as in
the canonical case~(\ref{eq:density-fixed-shape-start1}) [the
  difference in the lower limit of summation is irrelevant in this
  case since the summand vanishes very fast when $\hx_k$ differs from
  $x$]. Therefore, in the bulk, $n^{\text{gh}}(x)=n_b$ as expected.

When $x_m-x$ is of order $O(1/\sqrt{\alpha})$, once again the
exponential term in the denominator vanishes in the limit
$\alpha\to\infty$. The resulting expression is transformed into an
integral over the variable $\epsilon_{k,m}$, and following identical
calculations as the ones from
subsection~\ref{sec:density-canonic-fixed-shape}, we find
that, $n^{\text{gh}}(x)=n^{\text{hs}}(x)$, that is the same
result~(\ref{eq:density-border-xm}) as for the hard wall
boundary. This is somehow expected since, the boundary at $x=x_m$ is 
of the hard wall type. Notice that the density profile near this
boundary does not depend on the fugacity $\zeta$.

\begin{figure}
\begin{center}
\includegraphics[width=\GraphicsWidth]{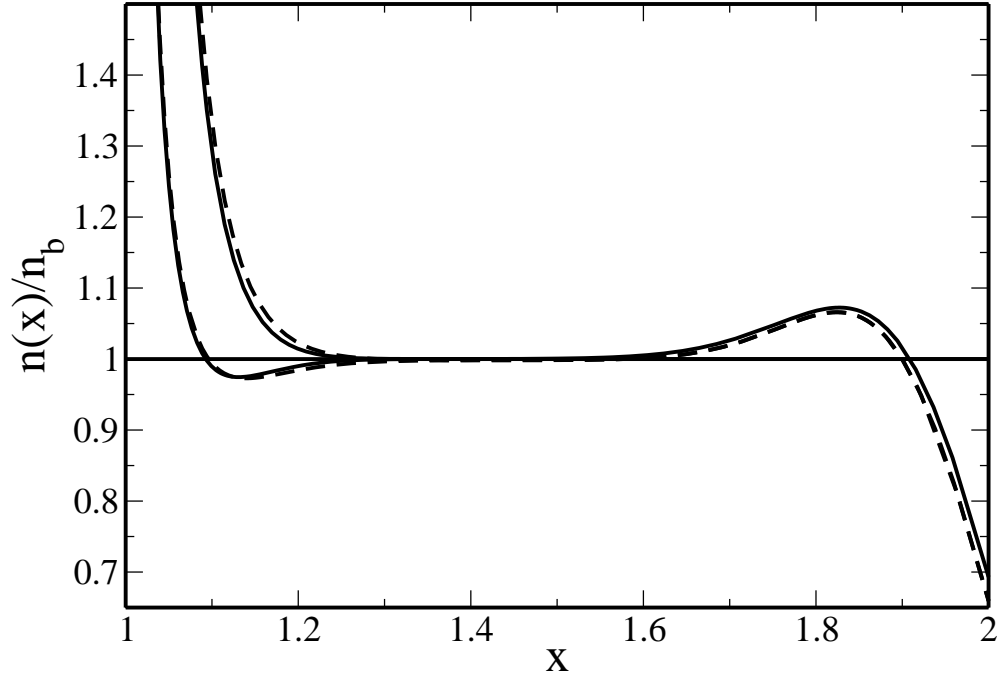}
\end{center}
\caption{The normalized one-body density $n^{\text{gh}}(x)/n_b$, in
  the grounded horizon case. The dashed lines correspond to a
  numerical evaluation, obtained from~(\ref{eq:densite-somme}), with
  $N=100$, $x_m=2$ and $\alpha=4.15493$ and truncating the sum to 301
  terms (the lower value of $k$ is $-200$). The gap close to the
  metallic boundary has been chosen equal to $w=0.01$. The full lines
  correspond to the asymptotic result in the fixed shape limit when
  $\alpha\to\infty$, and $x_m=2$ fixed. The two upper curves
  correspond to a fugacity given by
  $\zeta\sqrt{\alpha}/(2n_b)=\tilde{\zeta}L\sqrt{\pi/n_b}=1$, while
  the two lower ones correspond to $\tilde{\zeta}L\sqrt{\pi/n_b}=0.1$.
  Notice how the value of the fugacity only affects the density
  profile close to the metallic boundary $x=1$.  }
\label{fig:density-metal-fixed-shape}
\end{figure}

The last case is for the density profile close to the metallic
boundary, when $x-1$ is of order $O(1/\sqrt{\alpha})$. In this case,
contrary to the previous ones, the exponential term in the denominator
does not vanish. Expanding it around $\hx_k=1$, we have
\begin{equation}
  n^{\text{gh}}(x)=\zeta
  \sum_{k=-\infty}^{N_b}
  \frac{e^{-\alpha p'(x) (x-\hx_k)^2/x}}{
    e^{-\epsilon_{k,1}^2}
    +\frac{\zeta \sqrt{\alpha \pi \hx_k p'(\hx_k)}}{2 n_b}
    \left[ \erf(\tilde{\epsilon}_{k,1})+1
      \right]}
\,.
\end{equation}
Transforming the summation into an integral over the variable
$t=-\epsilon_{k,1}$, we find
\begin{equation}
  n^{\text{gh}}(x)=
  \zeta \sqrt{\alpha p'(1)}
  \int_{-\infty}^{+\infty}
  \frac{e^{-[t+\sqrt{\alpha p' (1)}(x-1)]^2}\ dt}{
    e^{-t^2}+\frac{\zeta\sqrt{\alpha \pi p'(1)}}{2 n_b}
    \erfc(t+\sqrt{2\pi n_b}W)}
  \,.
\end{equation}
For purposes of comparison with Ref.~\cite{Forrester85}, this can be
rewritten as
\begin{equation}
  n^{\text{gh}}(x)=
  \zeta \sqrt{\alpha p'(1)}
  e^{-\alpha p'(1)[(x-1-w)^2-w^2]}
  \int_{-\infty}^{+\infty}
  \frac{e^{-2\sqrt{\alpha p'(1)} (x-1)t}\ dt}{
    1+\frac{\zeta\sqrt{\alpha \pi p'(1)}}{2 n_b}
    \erfc(t)e^{(t-\sqrt{2\pi n_b}W)^2}}
  \,.
\end{equation}
Which is very similar to the density profile near a plane metallic
wall in flat space found in Ref.~\cite{Forrester85} [there is a
  small difference, due to the fact that in~\cite{Forrester85} the
  background did not extend up to the metallic wall, but also had a
  gap, contrary to our present
  model]. Fig.~\ref{fig:density-metal-fixed-shape} shows the density
profile for two different values of the fugacity, and compares the
asymptotic results with a direct numerical evaluation of the density.

Interestingly, one again, the density profile shows a universality
feature, in the sense that it is essentially the same as for a flat
space. As in the flat space, the fugacity controls the excess charge
which accumulates near the metallic wall. Only the density profile
close to the metallic wall depends on the fugacity. In the bulk, the
density is constant, equal to the background density. Close to the
other boundary (the hard wall one), the density profile is the same as
in the other models from previous sections, and it does not depend on
the fugacity.

\section{Conclusions}
\label{sec:conclusions}

The two-dimensional one-component classical plasma has been studied on
Flamm's paraboloid (the Riemannian surface obtained from the spatial
part of the Schwarzchild metric).  The one-component classical plasma
had long been used as the simplest microscopic model to describe many
Coulomb fluids such as electrolytes, plasmas, molten
salts~\cite{March84}. Recently it has also been studied on curved
surfaces as the cylinder, the sphere, and the pseudosphere. From this
point of view, this work presents new results as it describes the
properties of the plasma on a surface that had never been considered
before in this context.

The Coulomb potential on this surface has been carefully determined.
When we limit ourselves to study only the upper or lower half parts
($\mathcal{S}_{\pm}$) of the surface (see Fig.~\ref{fig:surf}) the
Coulomb potential is $G^{\text{hs}}(\qq,\qq')= -\ln |z-z'|+
\text{constant}$, with the appropriate set of coordinates
$(x,\varphi)$ defined in section~\ref{sec:good-coordinates}, and
$z=xe^{i\varphi}$. When charges from the upper part are allowed to
interact with particles from the lower part then the Coulomb potential
turns out to be $G^{\text{ws}}(\qq,\qq')=
-\ln(|z-z'|/\sqrt{|zz'|})+\text{constant}$. When the charges live in
the upper part with the horizon grounded, the Coulomb potential can be
determined using the method of images form electrostatics.

Since the Coulomb potential takes a form similar to the one of a flat
space, this allows to use the usual
techniques~\cite{Jancovici81b,Alastuey81} to compute the thermodynamic
properties when the coupling constant $\Gamma=\beta q^2=2$.

Two different thermodynamic limits have been considered: the one where
the radius $R$ of the ``disk'' confining the plasma is allowed to
become very big while keeping the surface hole radius $M$ constant,
and the one where both $R\to\infty$ and $M\to\infty$ with the ratio
$R/M$  kept constant (fixed shape limit). In both limits we computed
the free energy up to corrections of order $O(1)$.

The plasma on half surface is found to be thermodynamically stable, in
both types of thermodynamic limit, upon choosing the arbitrary
additive constant in the Coulomb potential equal to $-\ln
M+\text{constant}$. The system on the full surface is found to be
stable upon choosing the constant in the Coulomb potential equal to
$-\ln(M x_m) + \text{constant}$ where
$x_m=(\sqrt{R}+\sqrt{R-2M})^2/(2M)$.

In the limit $R\to\infty$ while keeping $M$ fixed, most of the surface
available to the particles is almost flat, therefore the bulk free
energy is the same as in flat space, but corrections from the flat
case, due to the curvature effects, appear in the terms proportional
to $R$ and the terms proportional to $\ln R$. These corrections are
different for each case (half or whole surface).

The asymptotic expansion at fixed shape ($\alpha\to\infty$) presents a
different value for the bulk free energy than in the flat space, due
to the curvature corrections. On the other hand, the perimeter
corrections to the free energy turn out to be the same as for a flat
space. This expansion of the free energy does not exhibit the
logarithmic correction, $\ln \alpha$, in agreement with the fact that
the Euler characteristic of this surface vanishes.

For completeness, we also studied the system on half surface letting
the particles interact through the Coulomb potential
$G^{\text{ws}}$. In this mixed case the result for the free energy is
simply one-half the one found for the system on the full surface.

In the case where the ``horizon'' is grounded (metallic boundary), the
system is studied in the grand canonical ensemble. The limit $R\to
\infty$ with $M$ fixed, reproduces the same results as the case of the
half surface with potential $G^{\text{hs}}$ up to $O(1)$ corrections,
because the effects of the size of the metallic boundary remain
$O(1)$. More interesting is the thermodynamic limit at fixed shape,
where we find that the bulk thermodynamics are the same as for the
half surface with potential $G^{\text{hs}}$, but a perimeter
correction associated to the metallic boundary appears. This turns out
to be the same as for a flat space. This perimeter correction
(``surface'' tension) $\beta \gamma_{\text{metal}}$ depends on the
value of the fugacity. In the grand canonical formalism, the system
can be nonneutral, in the bulk the system is locally neutral, and the
excess charge is found near the metallic boundary. In contrast, the
outer hard wall boundary (at $x=x_m$), exhibits the same density
profile as in the other cases, independent of the value of the
fugacity. This reflects in a perimeter contribution $\beta
\gamma_{\text{hard}}$ equal to the one of the previous cases.

The plasma on Flamm' s paraboloid is not homogeneous due to the fact
that the curvature of the surface is not constant. When the horizon
shrinks to a point the upper half surface reduces to a plane and one
recovers the well known result valid for the one component plasma on
the plane.  In the same limit the whole surface reduces to two flat
planes connected by a hole at the origin. 

We carefully studied the one body density for several different
situations: plasma on half surface with potential $G^{\text{hs}}$ and
$G^{\text{ws}}$, plasma on the whole surface with potential
$G^{\text{ws}}$, and plasma on half surface with the horizon
grounded. When only one-half of the surface is occupied by the plasma,
if we use $G^{\text{hs}}$ as the Coulomb potential, the density shows
a peak in the neighborhoods of each boundary, tends to a finite
value at the boundary and to the background density far from it, in the
bulk. If we use $G^{\text{ws}}$, instead, the qualitative behavior of the
density remains the same. In the thermodynamic limit at fixed shape,
we find that the density profile is the same as in flat space near a
hard wall, regardless of the Coulomb potential used.

In the grounded horizon case the density reaches the background
density far from the boundaries. In this case, the fugacity and the
background density control the density profile close to the metallic
boundary (horizon). In the bulk and close to the outer hard wall
boundary, the density profile is independent of the fugacity. In the
thermodynamic limit at fixed shape, the density profile is the same as
for a flat space.

Internal and external screening sum rules have been briefly
discussed. Nevertheless, we think that systems with non constant
curvature should deserve a revisiting of all the common sum rules for
charged fluids.

\begin{acknowledgments}

Riccardo Fantoni would like to acknowledge the support from the italian
MIUR (PRIN-COFIN 2006/2007). He would also wish to dedicate this
work to his wife Ilaria Tognoni who is undergoing a very delicate and
reflexive period of her life.

G.~T.~acknowledges partial financial support from Comit\'e de
Investigaciones y Posgrados, Facultad de Ciencias, Universidad de los
Andes.

\end{acknowledgments}

\appendix

\section{Green function of Laplace equation}
\label{app:green}

In this appendix, we illustrate the calculation of the Green function
using the original system of coordinates $(r,\varphi)$. The Coulomb
potential generated at $\qq=(r,\varphi)$ by a unit charge placed at
$\qq_0=(r_0,\varphi_0)$ with $r_0>2M$ satisfies the
Poisson equation
\begin{eqnarray}
\Delta G(r,\varphi;r_0,\varphi_0) = 
-2\pi\delta(r-r_0)\delta(\varphi-\varphi_0)/\sqrt{g}~,
\end{eqnarray}
where $g=\det (g_{\mu\nu})=r^2/(1-2M/r)$. To solve this equation, we
expand the Green function $G$ and the second delta distribution in a
Fourier series as follows
\begin{eqnarray} \label{gexp}
G(r,\varphi;r_0,\varphi_0)&=&
\sum_{n=-\infty}^\infty e^{in(\varphi-\varphi_0)} g_n(r,r_0)~,\\
\delta(\varphi-\varphi_0)&=&\frac{1}{2\pi}
\sum_{n=-\infty}^\infty e^{in(\varphi-\varphi_0)}~,
\end{eqnarray}
to obtain an ordinary differential equation for $g_n$
\begin{eqnarray}
\left[\left(1-\frac{2M}{r}\right)\frac{\partial^2}{\partial r^2}
+\left(\frac{1}{r}-\frac{M}{r^2}\right)\frac{\partial}{\partial
r}-\frac{n^2}{r^2}\right]g_n(r,r_0)=-\delta(r-r_0)/\sqrt{g}~.
\end{eqnarray}
To solve this equation we first solve the homogeneous one for $r<r_0$:
$g_{n,-}(r,r_0)$ and $r>r_0$: $g_{n,+}(r,r_0)$. The solution is, for
$n\neq 0$,
\begin{eqnarray}
g_{n,\pm}(r,r_0)=A_{n,\pm}(\sqrt{r}+\sqrt{r-2M})^{2n}+
B_{n,\pm}(\sqrt{r}+\sqrt{r-2M})^{-2n}~,
\end{eqnarray} 
and, for $n=0$, one finds
\begin{eqnarray}
g_{0,\pm}(r,r_0)=A_{0,\pm}+B_{0,\pm}\ln(\sqrt{r}+\sqrt{r-2M})~.
\end{eqnarray}
The form of the solution immediately suggest that it is more
convenient to work with the variable
$x=(\sqrt{r}+\sqrt{r-2M})^{2}/(2M)$. For this reason, we introduced
this new system of coordinates $(x,\varphi)$ which is used in the main
text.

\section{Asymptotic expansions of $\mathcal{B}_N(k)$,
  $\tilde{\mathcal{B}}_N(k)$ and $\hat{\mathcal{B}}_N(k)$}
\label{app:gamma}

\subsection{Asymptotic expansion of $\mathcal{B}_N(k)$}

\subsubsection{Limit $N\to\infty$, $x_m\to\infty$, and fixed $\alpha$}

Doing the change of variable $s=\alpha p(x)$ in the
integral~(\ref{gamma}), we have
\begin{equation}
  \mathcal{B}_N(k)=\frac{1}{n_b} \int_0^N x^{2k} 
  e^{-\alpha h(x)}\, ds
\end{equation}
where $x$ is related to the variable of integration $s$ by $s=\alpha
p(x)$. The limit $k\to\infty$ and $N\to\infty$ can be obtained using
Laplace method~\cite{Bender99}. To this end, let us write
$\mathcal{B}_N(k)$ as
\begin{equation}
  \mathcal{B}_N(k)=\frac{k}{n}\int_0^{N/k}
  e^{k\phi_k(t)}\,dt
\end{equation}
where we made the change of variable $t=s/k$ and we defined
\begin{equation}
  \label{eq:phi-k}
  \phi_{k}(t)= 2 \ln x -\frac{\alpha}{k}\, h(x)
\end{equation}
where 
\begin{equation}
  \label{eq:def-x}
  x=p^{-1}(kt/\alpha)
  \,.
\end{equation}
The derivative of $\phi_k$ is
\begin{eqnarray}
  \phi_k'(t)&=&
  \frac{1}{x}\frac{dx}{dt} (1-t)\\
  &=& \frac{2k}{\alpha x p' (x)}(1-t)
\end{eqnarray}
where we have used the definition~(\ref{eq:def-x}) of $x$ and the
properties~(\ref{eq:p-h-deriv}) of $h$ and $p$.

The maximum of $\phi_k(t)$ is obtained when $t=1$. At this point we
have
\begin{subequations}
\begin{eqnarray}
  \phi_k''(1)&=&-\frac{2 k}{\alpha \hx_k p' (\hx_k)}
  =-1+O(1/\sqrt{k})
  \\
  \phi_k^{(3)}(1)&=&\frac{4 k^2}{\alpha^2}
  \frac{p'(\hx_k)+x_k p''(\hx_k)}{\hx_k^2 p'(\hx_k)^3}
  =2+O(1/\sqrt{k})\\
  \phi_k^{(4)}(1)&=&\frac{6k^3}{\alpha^3 p' (\hx_k)}
  \frac{d}{dx}\left[
  \frac{p'(x)+x p''(x)}{x^2 p'(x)^3}
  \right]_{x=\hx_k}
  =-6+O(1/\sqrt{k})
\end{eqnarray}
\label{eq:phis-k}
\end{subequations}
where 
\begin{equation}
  \hx_k=p^{-1}(k/\alpha)
  \,.
\end{equation}
Expanding $\phi_k(t)$ up to order $(t-1)^4$, and defining
$v=\sqrt{k|\phi''_k(1)|}\,(t-1)$, we have
\begin{eqnarray}
  \mathcal{B}_N(k)&=&
  \frac{\sqrt{k}e^{k\phi_k(1)}}{n \sqrt{|\phi''_k(1)|}}\,
  \int_{-\sqrt{k|\phi''_k(1)|}}^{(N-k)\sqrt{|\phi''_k(1)|/k}}
  e^{-v^2/2}
  \nonumber\\
  &&
  \times
  \left[
    1+\frac{v^3 \phi^{(3)}_{k}(1)}{3!\sqrt{k}|\phi''_k(1)|^{3/2}}
    +\frac{v^4 \phi_k^{(4)}(1)}{4! k |\phi''_k(1)|^2}
    +\frac{v^6 [\phi^{(3)}_{k}(1)]^2}{3!^2 2k |\phi''_k(1)|^3}
      +o\left(\frac{1}{k}\right)
    \right]
    \,dv\,.
\end{eqnarray}
Let us define
\begin{equation}
  \label{eq:epsilon-k}
  \epsilon_k=\sqrt{|\phi''_k(1)|}\,\frac{N-k}{\sqrt{2k}}
  =\frac{N-k}{\sqrt{2N}}+O(1/\sqrt{N})
\end{equation}
which is an order one parameter, since we are interested in an
expansion for $N$ and $k$ large with $N-k$ of order $\sqrt{N}$. Using
the integrals
\begin{eqnarray}
  \int_{-\infty}^{\epsilon} e^{-v^2/2}\,dv 
  & = &
  \sqrt{\frac{\pi}{2}}
  \left[ 1 + \erf\left(\frac{\epsilon}{\sqrt{2}}\right) \right]
  \\
  \int_{-\infty}^{\epsilon} e^{-v^2/2}\, v^3 \,dv 
  & = & 
  -(2+\epsilon^2) e^{-\epsilon^2/2}
  \\
  \int_{-\infty}^{\epsilon} e^{-v^2/2}\, v^4 \,dv 
  & = & 
  3
  \sqrt{\frac{\pi}{2}}
  \left[ 1 + \erf\left(\frac{\epsilon}{\sqrt{2}}\right) \right]
  -e^{-\epsilon^2/2}\epsilon(3+\epsilon^2)
  \\
  \int_{-\infty}^{\epsilon} e^{-v^2/2}\, v^6 \,dv 
  & = & 
  15
  \sqrt{\frac{\pi}{2}}
  \left[ 1 + \erf\left(\frac{\epsilon}{\sqrt{2}}\right) \right]
  -e^{-\epsilon^2/2}\epsilon(15+5\epsilon^2+\epsilon^4)
\end{eqnarray}
where $\erf(z)=(2/\sqrt{\pi}) \int_0^{z} e^{-u^2}\,du$ is the error
function, we find in the limit $N\to\infty$, $k\to\infty$, and finite
$\epsilon_k$,
\begin{eqnarray}
  \label{eq:asymptics-B}
  \mathcal{B}_N(k)&=&
  \sqrt{\frac{\pi k}{2 |\phi''_k(1)|}}
  \frac{e^{k\phi_k(1)}}{n}
  \left[1+\erf\left({\epsilon_k}\right)\right]
  \left[1+\frac{1}{12 k} + \frac{1}{\sqrt{k}}\,\xi_1(\epsilon_k)
    +\frac{1}{k}\,\xi_2(\epsilon_k)\right]
  \,.
\end{eqnarray}
The functions $\xi_1(\epsilon_k)$ and $\xi_2(\epsilon_k)$ contain 
terms proportional $e^{-\epsilon_k^2}$, from the Gaussian integrals
above. However, as explained in the main text, these do not contribute
to the final result for the partition function up to order $O(1)$,
because the exponential term $e^{-\epsilon_k^2}$ make convergent and
finite the integrals of these functions that appear in the
calculations, giving terms of order $O(1)$ and $O(1/\sqrt{N})$
respectively.

\subsubsection{Limit $N\to\infty$, $\alpha\to\infty$, fixed $x_m$}

For the determination of the thermodynamic limit at fixed shape, we
also need the asymptotic behavior of $\mathcal{B}_N(k)$ when
$\alpha\to\infty$ at fixed $x_m$. We write $\mathcal{B}_N(k)$ as
\begin{equation}
  \mathcal{B}_N(k)=\frac{\alpha}{n_b}
  \int_{1}^{x_m} e^{-\alpha [ h(x)-2p(\hx_k) \ln x ]} \, p'(x)\,dx
  \,,
\end{equation}
where we have defined once again $\hx_k$ by $k=\alpha p(\hx_k)$. We apply
Laplace method for $\alpha\to\infty$. Let
\begin{equation}
  F(x)=h(x)-2p(\hx_k)\ln x\,.
\end{equation}
$F$ has a minimum for $x=\hx_k$ with
$F''(\hx_k)=2p'(\hx_k)/\hx_k$. Expanding to the order $(x-\hx_k)^2$ the
argument of the exponential and following calculations similar to the
ones of the previous section, we find
\begin{eqnarray}
  \mathcal{B}_N(k)
  &=&
  \frac{\sqrt{\alpha\pi \hx_k p'(\hx_k)}}{2n_b}
  e^{-\alpha [h(\hx_k)-2p(\hx_k) \ln \hx_k]}
  \left[
  \erf(\epsilon_{k,1})
  +\erf(\epsilon_{k,m})
    \right]
  \nonumber\\
  &&
  \times
    \left(1+ \frac{1}{\alpha} \xi_0(\hx_k) 
    + \frac{1}{\sqrt{\alpha}}
    \left[\xi_{1,m}(\epsilon_{k,m})
    +
    \xi_{1,1}(\epsilon_{k,1})
    \right]
    \right)
  \label{eq:BN-fixed-shape}
\end{eqnarray}
where
\begin{eqnarray}
  \label{eq:epsilon_km}
  \epsilon_{k,m} &=&\sqrt{\frac{ \alpha p'(x_m)}{x_m}}(x_m-\hx_k)
  \,,
  \\
  \label{eq:epsilon_k1}
  \epsilon_{k,1} &=&\sqrt{ \alpha p'(1)}(\hx_k-1)
  \,.
\end{eqnarray}
The terms with the error functions come from incomplete Gaussian
integral and take into account the contribution of values of $k$ such
that $x_m-\hx_k$ (or $\hx_k-1$) is of order $1/\sqrt{\alpha}$, or
equivalently $N-k$ (or $k$) of order $\sqrt{N}$.

The functions $\xi_0(\hx_k)$, $\xi_{1,1}(\epsilon_{k,1})$, and
$\xi_{1,m}(\epsilon_{k,m})$ can be computed explicitly, pushing the
expansion one order further. These next order corrections are
different than in the previous section, in particular $(1/\alpha)
\xi_0(\hx_k)\neq 1/(12 k)$.

However, these next order terms are not needed in the computation of
the partition function at order $O(1)$, since they give contributions
of order $O(1)$. Note in particular that the term $\xi_0(\hx_k)/\alpha$ gives
contributions of order $O(1)$, contrary to the previous limit studied
earlier where it gave contributions of order $\ln N$. Indeed, in the
logarithm of the partition function, this term gives a contribution
\begin{equation}
  \sum_{k=0}^{N} \frac{\xi_0(\hx_k)}{\alpha }
  = \frac{1}{\alpha}
  \int_{1}^{x_m} \alpha p'(x) \xi_0(x) \, dx + o(1) = O(1)
\,.
\end{equation}

\subsection{Asymptotic expansions of $\tilde{\mathcal{B}}_N(k)$ and
  $\hat{\mathcal{B}}_N(k)$}

To study $\tilde{\mathcal{B}}_N(k)$, it is convenient to define
$k'=k-\frac{N}{2}$, then
\begin{equation}
  \tilde{\mathcal{B}}_N(k)=\frac{\alpha}{n_b}
  \int_{1/x_m}^{x_m} x^{2k'} e^{-\alpha h(x)}\,x\,p'(x)\,dx
  \,,
\end{equation}
which is very similar to 
\begin{equation}
  \hat{\mathcal{B}}_N(k)=\frac{\alpha}{n_b}
  \int_{1}^{x_m} x^{2k} e^{-\alpha h(x)}\,x\,p'(x)\,dx
  \,.
\end{equation}
changing $k'$ by $k$, and taking into account the extended domain of
integration $[1/x_m,1]$ for $\tilde{\mathcal{B}}_N$. As in the
previous section, the asymptotic expansions for
$\tilde{\mathcal{B}}_N(k)$ and $\hat{\mathcal{B}}_N(k)$ can be
obtained using Laplace method. Notice that for
$\tilde{\mathcal{B}}_N(k)$, $k'$ is in the range
$[-\frac{N}{2},\frac{N}{2}]$. When $k'<0$, the maximum of the
integrand is in the region $[1/x_m,1]$, and when $k'>0$, the maximum
is in the region $[1,x_m]$. Due to the fact that the contribution to
the integral from the region $[1/x_m,1]$ is negligible when $k'>0$,
the asymptotics for $\hat{\mathcal{B}}_N(k)$ will be the same as those
for $\tilde{\mathcal{B}}_N(k)$, for $k'>0$, doing the change $k\to
k'$. Therefore, we present only the derivation of the asymptotics of
$\tilde{\mathcal{B}}_N$.

\subsubsection{Limit $N\to\infty$, $x_m\to\infty$, and fixed $\alpha$}

We proceed as for ${\mathcal{B}}_N(k)$, defining the variable of
integration $t=\alpha p(x)/k'$, then
\begin{equation}
  \tilde{\mathcal{B}}_N(k)=\frac{|k'|}{n_b}
  \int_{-\frac{N}{2|k'|}}^{\frac{N}{2|k'|}}
  x\,e^{k'\phi_{k'}(t)}\,dt
\end{equation}
where $\phi_{k'}(t)$ is the same function defined in
equation~(\ref{eq:phi-k}). Now we apply Laplace method to compute this
integral. The main difference with the calculations done for
$\mathcal{B}_{N}$ are the following. First, taking into account that $k'$ can
be positive or negative, we should note that
\begin{eqnarray}
  \phi_{k'}''(1)&=&
  \begin{cases}
  -1+O(1/\sqrt{|k'|})& k'>0    \\
  1+O(1/\sqrt{|k'|})& k'<0    
  \end{cases}
  \\
  \phi_{k'}^{(3)}(1)&=&
  \begin{cases}
    2+O(1/\sqrt{|k'|})& k'>0\\
    -2+O(1/\sqrt{|k'|})& k'<0
  \end{cases}
  \\
  \phi_{k'}^{(4)}(1)
  &=&
  \begin{cases}
  -6+O(1/\sqrt{|k'|})&k'>0\\
  6+O(1/\sqrt{|k'|})&k'<0
  \end{cases}
\end{eqnarray}
Second, we also need to expand $x$ close to the maximum which is
obtained for $t=1$,
\begin{equation}
  x=\hx_{k'}[1+a (t-1) + b(t-1)^2 +O((t-1)^3)]
\end{equation}
with
\begin{equation}
  a=\frac{p(\hx_{k'})}{\hx_{k'} p'(\hx_{k'})}
  =
  \begin{cases}
    \frac{1}{2}+O(1/\sqrt{|k'|})&k'>0\\
    -\frac{1}{2}+O(1/\sqrt{|k'|})&k'<0
  \end{cases}
\end{equation}
and
\begin{equation}
  b=-\frac{p(\hx_{k'})^2 p''(\hx_{k'})}{2 \hx_{k'} p'(\hx_{k'})^3}
  =
  \begin{cases}
    -\frac{1}{8}+O(1/\sqrt{|k'|})& k'>0\\
    \frac{3}{8}+O(1/\sqrt{|k'|})& k'<0
  \end{cases}
\end{equation}
Notice in particular that for the term $b$, the difference between
positive and negative values of $k'$ is not only a change of
sign. This is to be expected since the function $x$ is not invariant
under the change $x\to 1/x$. 

Following very similar calculations to the ones done for
$\mathcal{B}_N$ with the appropriate changes mentioned above, we
finally find
\begin{eqnarray}
  \label{eq:asympt-tildeBN}
  \tilde{B}_N(k)&=&\frac{\hx_{k'}}{2n_b}\sqrt{\pi\alpha \hx_{k'} 
    p'(\hx_{k'})}
  e^{-\alpha [h(\hx_{k'})-2p(\hx_{k'}) \ln \hx_{k'}]}
  \nonumber\\
  &&\times
  \left[\erf\left(\epsilon_{k,\min}\right)+\erf\left(\epsilon_{k,\max}\right)
    \right]
  \left[1+\left(\frac{1}{12}+ c \right)\frac{1}{|k'|}+\cdots\right]
\end{eqnarray}
with
\begin{equation}
  c=
  \begin{cases}
    \frac{3}{8}&k'>0\\
    -\frac{1}{8}&k'<0
  \end{cases}
\end{equation}
and
\begin{subequations}
\label{eq:epsilons-min-max}
\begin{eqnarray}
  \epsilon_{k,\max} &=&\sqrt{\frac{ \alpha p'(x_m)}{x_m}}
  (x_m-\hx_{k-\frac{N}{2}})
  \,,
  \\
  \epsilon_{k,\min} &=&\sqrt{\frac{ \alpha p'(1/x_m)
    }{1/x_m}}\left(\hx_{k-\frac{N}{2}}-\frac{1}{x_m}\right)
  \,.
\end{eqnarray}
\end{subequations}
The dots in~(\ref{eq:asympt-tildeBN}) represent contributions of lower
order and of functions of $\epsilon_{k,\min}$ and $\epsilon_{k,\max}$
that give $O(1)$ contributions to the partition function. Comparing to
the asymptotics of $\mathcal{B}_N$ we notice two differences: the
factor $\hx_{k'}$ multiplying all the expressions and the correction
$c/|k'|$.

\subsubsection{Limit $N\to\infty$, $\alpha\to\infty$, and fixed $x_m$}

The asymptotic expansion of $\tilde{\mathcal{B}}_N$ in this fixed shape
situation is simpler, since we do not need the terms of order
$1/\alpha$. Doing similar calculations as the ones done for
$\mathcal{B}_N$ taking into account the additional factor $x$ in the
integral we find
\begin{eqnarray}
  \tilde{\mathcal{B}}_N(k)
  &=&
  \frac{\hx_{k'}\sqrt{\alpha\pi \hx_{k'} p'(\hx_{k'})}}{2n_b}
  e^{-\alpha [h(\hx_{k'})-2p(\hx_{k'}) \ln \hx_{k'}]}
  \left[
  \erf(\epsilon_{k,\min})
  +\erf(\epsilon_{k,\max})
    \right]\,.
  \label{eq:tildeBN-fixed-shape}
\end{eqnarray}

\bibliographystyle{apsrev}
\bibliography{2docp}

\begin{thebibliography}{28}
\expandafter\ifx\csname natexlab\endcsname\relax\def\natexlab#1{#1}\fi
\expandafter\ifx\csname bibnamefont\endcsname\relax
  \def\bibnamefont#1{#1}\fi
\expandafter\ifx\csname bibfnamefont\endcsname\relax
  \def\bibfnamefont#1{#1}\fi
\expandafter\ifx\csname citenamefont\endcsname\relax
  \def\citenamefont#1{#1}\fi
\expandafter\ifx\csname url\endcsname\relax
  \def\url#1{\texttt{#1}}\fi
\expandafter\ifx\csname urlprefix\endcsname\relax\def\urlprefix{URL }\fi
\providecommand{\bibinfo}[2]{#2}
\providecommand{\eprint}[2][]{\url{#2}}

\bibitem[{\citenamefont{{S. F. Edwards and A. Lenard}}(1962)}]{Edwards62}
\bibinfo{author}{\bibnamefont{{S. F. Edwards and A. Lenard}}},
  \bibinfo{journal}{J. Math. Phys.} \textbf{\bibinfo{volume}{{\bf 3}}},
  \bibinfo{pages}{778} (\bibinfo{year}{1962}).

\bibitem[{\citenamefont{Jancovici}(1981)}]{Jancovici81b}
\bibinfo{author}{\bibfnamefont{B.}~\bibnamefont{Jancovici}},
  \bibinfo{journal}{Phys. Rev. Lett.} \textbf{\bibinfo{volume}{{\bf 46}}},
  \bibinfo{pages}{386} (\bibinfo{year}{1981}).

\bibitem[{\citenamefont{{A. Alastuey and B. Jancovici}}(1981)}]{Alastuey81}
\bibinfo{author}{\bibnamefont{{A. Alastuey and B. Jancovici}}},
  \bibinfo{journal}{J. Phys. (France)} \textbf{\bibinfo{volume}{{\bf 42}}},
  \bibinfo{pages}{1} (\bibinfo{year}{1981}).

\bibitem[{\citenamefont{{B. Jancovici and G. T\'{e}llez}}(1996)}]{Jancovici96}
\bibinfo{author}{\bibnamefont{{B. Jancovici and G. T\'{e}llez}}},
  \bibinfo{journal}{J. Stat. Phys.} \textbf{\bibinfo{volume}{{\bf 82}}},
  \bibinfo{pages}{609} (\bibinfo{year}{1996}).

\bibitem[{\citenamefont{{B. Jancovici, G. Manificat, and C.
  Pisani}}(1994)}]{Jancovici94}
\bibinfo{author}{\bibnamefont{{B. Jancovici, G. Manificat, and C. Pisani}}},
  \bibinfo{journal}{J. Stat. Phys.} \textbf{\bibinfo{volume}{{\bf 76}}},
  \bibinfo{pages}{307} (\bibinfo{year}{1994}).

\bibitem[{\citenamefont{{M. L. Rosinberg, L. Blum}}(1984)}]{Rosinberg84}
\bibinfo{author}{\bibnamefont{{M. L. Rosinberg, L. Blum}}},
  \bibinfo{journal}{J. Chem. Phys.} \textbf{\bibinfo{volume}{{\bf 81}}},
  \bibinfo{pages}{3700} (\bibinfo{year}{1984}).

\bibitem[{\citenamefont{{Ph. Choquard}}(1981)}]{Choquard81}
\bibinfo{author}{\bibnamefont{{Ph. Choquard}}}, \bibinfo{journal}{Helv. Phys.
  Acta} \textbf{\bibinfo{volume}{{\bf 54}}}, \bibinfo{pages}{332}
  (\bibinfo{year}{1981}).

\bibitem[{\citenamefont{{Ph. Choquard, P. J. Forrester, and E. R.
  Smith}}(1983)}]{Choquard83}
\bibinfo{author}{\bibnamefont{{Ph. Choquard, P. J. Forrester, and E. R.
  Smith}}}, \bibinfo{journal}{J.~Stat.~Phys.} \textbf{\bibinfo{volume}{{\bf
  33}}}, \bibinfo{pages}{13} (\bibinfo{year}{1983}).

\bibitem[{\citenamefont{Caillol}(1981)}]{Caillol81}
\bibinfo{author}{\bibfnamefont{J.~M.} \bibnamefont{Caillol}},
  \bibinfo{journal}{J. Phys. (Paris) -- Lett.} \textbf{\bibinfo{volume}{{\bf
  42}}}, \bibinfo{pages}{L} (\bibinfo{year}{1981}).

\bibitem[{\citenamefont{{P. J. Forrester, B. Jancovici, and J.
  Madore}}(1992)}]{Jancovici92}
\bibinfo{author}{\bibnamefont{{P. J. Forrester, B. Jancovici, and J. Madore}}},
  \bibinfo{journal}{J. Stat. Phys.} \textbf{\bibinfo{volume}{{\bf 69}}},
  \bibinfo{pages}{179} (\bibinfo{year}{1992}).

\bibitem[{\citenamefont{{P. J. Forrester and B.
  Jancovici}}(1996)}]{Jancovici96b}
\bibinfo{author}{\bibnamefont{{P. J. Forrester and B. Jancovici}}},
  \bibinfo{journal}{J. Stat. Phys.} \textbf{\bibinfo{volume}{{\bf 84}}},
  \bibinfo{pages}{337} (\bibinfo{year}{1996}).

\bibitem[{\citenamefont{{G. T\'ellez and P. J. Forrester}}(1999)}]{Tellez99}
\bibinfo{author}{\bibnamefont{{G. T\'ellez and P. J. Forrester}}},
  \bibinfo{journal}{J. Stat. Phys.} \textbf{\bibinfo{volume}{{\bf 97}}},
  \bibinfo{pages}{489} (\bibinfo{year}{1999}).

\bibitem[{\citenamefont{Jancovici}(2000)}]{Jancovici00}
\bibinfo{author}{\bibfnamefont{B.}~\bibnamefont{Jancovici}},
  \bibinfo{journal}{J. Stat. Phys.} \textbf{\bibinfo{volume}{{\bf 99}}},
  \bibinfo{pages}{1281} (\bibinfo{year}{2000}).

\bibitem[{\citenamefont{{B. Jancovici and G. T\'{e}llez}}(1998)}]{Jancovici98}
\bibinfo{author}{\bibnamefont{{B. Jancovici and G. T\'{e}llez}}},
  \bibinfo{journal}{J. Stat. Phys.} \textbf{\bibinfo{volume}{{\bf 91}}},
  \bibinfo{pages}{953} (\bibinfo{year}{1998}).

\bibitem[{\citenamefont{{R. Fantoni, B. Jancovici, and G.
  T\'{e}llez}}(2003)}]{Fantoni03jsp}
\bibinfo{author}{\bibnamefont{{R. Fantoni, B. Jancovici, and G. T\'{e}llez}}},
  \bibinfo{journal}{J. Stat. Phys.} \textbf{\bibinfo{volume}{{\bf 112}}},
  \bibinfo{pages}{27} (\bibinfo{year}{2003}).

\bibitem[{\citenamefont{{B. Jancovici and G. T\'ellez}}(2004)}]{Jancovici04}
\bibinfo{author}{\bibnamefont{{B. Jancovici and G. T\'ellez}}},
  \bibinfo{journal}{J. Stat. Phys.} \textbf{\bibinfo{volume}{{\bf 116}}},
  \bibinfo{pages}{205} (\bibinfo{year}{2004}).

\bibitem[{\citenamefont{Kaniadakis}(2002)}]{kaniadakis02}
\bibinfo{author}{\bibfnamefont{G.}~\bibnamefont{Kaniadakis}},
  \bibinfo{journal}{Phys. Rev. E} \textbf{\bibinfo{volume}{{\bf 66}}},
  \bibinfo{pages}{056125} (\bibinfo{year}{2002}).

\bibitem[{\citenamefont{Kaniadakis}(2005)}]{kaniadakis05}
\bibinfo{author}{\bibfnamefont{G.}~\bibnamefont{Kaniadakis}},
  \bibinfo{journal}{Phys. Rev. E} \textbf{\bibinfo{volume}{{\bf 72}}},
  \bibinfo{pages}{036108} (\bibinfo{year}{2005}).

\bibitem[{\citenamefont{Ginibre}(1965)}]{Ginibre65}
\bibinfo{author}{\bibfnamefont{J.}~\bibnamefont{Ginibre}}, \bibinfo{journal}{J.
  Math. Phys.} \textbf{\bibinfo{volume}{{\bf 6}}}, \bibinfo{pages}{440}
  (\bibinfo{year}{1965}).

\bibitem[{\citenamefont{{M.~L.~Mehta}}(1991)}]{Mehta91}
\bibinfo{author}{\bibnamefont{{M.~L.~Mehta}}}, \emph{\bibinfo{title}{Random
  Matrices}} (\bibinfo{publisher}{Academic Press}, \bibinfo{year}{1991}).

\bibitem[{\citenamefont{{M. Abramowitz and A. Stegun}}(1965)}]{Abramowitz}
\bibinfo{author}{\bibnamefont{{M. Abramowitz and A. Stegun}}},
  \emph{\bibinfo{title}{{"Handbook of mathematical functions"}}}
  (\bibinfo{publisher}{Dover}, \bibinfo{address}{New York},
  \bibinfo{year}{1965}).

\bibitem[{\citenamefont{{R.~Wong}}(1989)}]{Wong89}
\bibinfo{author}{\bibnamefont{{R.~Wong}}}, \emph{\bibinfo{title}{Asymptotic
  Approximations of Integrals}} (\bibinfo{publisher}{Academic Press},
  \bibinfo{year}{1989}).

\bibitem[{\citenamefont{{B. Jancovici}}(1982)}]{Jancovici82}
\bibinfo{author}{\bibnamefont{{B. Jancovici}}}, \bibinfo{journal}{J. Stat.
  Phys.} \textbf{\bibinfo{volume}{{\bf 28}}}, \bibinfo{pages}{43}
  (\bibinfo{year}{1982}).

\bibitem[{\citenamefont{{Ph. A. Martin}}(1988)}]{Martin88}
\bibinfo{author}{\bibnamefont{{Ph. A. Martin}}}, \bibinfo{journal}{Rev. Mod.
  Phys.} \textbf{\bibinfo{volume}{{\bf 60}}}, \bibinfo{pages}{1075}
  (\bibinfo{year}{1988}).

\bibitem[{\citenamefont{{P.~J.~Forrester}}(1985)}]{Forrester85}
\bibinfo{author}{\bibnamefont{{P.~J.~Forrester}}},
  \bibinfo{journal}{J.~Phys.~A: Math.~Gen.} \textbf{\bibinfo{volume}{{\bf
  18}}}, \bibinfo{pages}{1419} (\bibinfo{year}{1985}).

\bibitem[{\citenamefont{Zinn-Justin}(1993)}]{ZinnJustin}
\bibinfo{author}{\bibfnamefont{J.}~\bibnamefont{Zinn-Justin}},
  \emph{\bibinfo{title}{{"Quantum Field Theory and Critical Phenomena"}}}
  (\bibinfo{publisher}{Clarendon Press}, \bibinfo{address}{Oxford},
  \bibinfo{year}{1993}), \bibinfo{edition}{2nd} ed.

\bibitem[{\citenamefont{{N. H. March, and M. P. Tosi}}(1984)}]{March84}
\bibinfo{author}{\bibnamefont{{N. H. March, and M. P. Tosi}}},
  \emph{\bibinfo{title}{{"Coulomb liquids"}}} (\bibinfo{publisher}{Academic
  Press}, \bibinfo{year}{1984}).

\bibitem[{\citenamefont{{C.~M.~Bender, and S.~ A.~ Orzag}}(1999)}]{Bender99}
\bibinfo{author}{\bibnamefont{{C.~M.~Bender, and S.~ A.~ Orzag}}},
  \emph{\bibinfo{title}{Advanced Mathematical Methods for Scientists and
  Engineers: Asymptotic Methods and Perturbation Theory}}
  (\bibinfo{publisher}{Springer}, \bibinfo{year}{1999}).

\end{thebibliography}

\end{document}